%% file: main.tex
\documentclass[runningheads]{llncs}
\input{preamble}

\begin{document}
	\setdefaultenum{(i)}{(a)}{(a)}{(a)}
	\moreless{}{
		\setlength{\abovedisplayskip}{.6em}
		\setlength{\belowdisplayskip}{.6em}
	}
	
	\title{%
		Model-based Fault Classification \texorpdfstring{\newline}{ }for Automotive Software
	}
	\titlerunning{%
		Model-based Fault Classification for Automotive Software
	}

	\author{%
		Mike Becker\inst{1}
		\and
		Roland Meyer\inst{1}
		\and
		Tobias Runge\inst{1,2}
		\and\\
		Ina Schaefer\inst{1,2}
		\and
		Sören van der Wall\inst{1}
		\and
		Sebastian Wolff\inst{3}
	}
	\authorrunning{M. Becker et al.}
	\institute{%
		TU Braunschweig, Germany,
		\email{\{mike.becker,roland.meyer,s.van-der-wall\}@tu-bs.de}
		\and
		KIT Karlsruhe, Germany,
		\email{\{tobias.runge,ina.schaefer\}@kit.edu}
		\and
		New York University, USA,
		\email{sebastian.wolff@nyu.edu}
	}

	\maketitle

	\input{content/abstract}
	\input{content/base}

	\subsubsection{Acknowledgements}
	The results were obtained in the projects \emph{``Virtual Test Analyzer I -- III''}, conducted in collaboration with \emph{IAV GmbH}.
	The last author is supported by a Junior Fellowship from the Simons Foundation (855328, SW).

	\bibliographystyle{splncs04}
	\bibliography{bib_manual}

	\moreless{
		\appendix
		\include{appendix/base}
	}{}
\end{document}

%% file: preamble.tex
\usepackage[utf8]{inputenc}
\usepackage[T1]{fontenc}
\usepackage[american]{babel}
\usepackage{amssymb}
\usepackage{stmaryrd}
\usepackage{mathtools}
\usepackage{galois}
\usepackage{xcolor}
\usepackage{paralist}
\usepackage{enumitem}
\usepackage{xspace}
\usepackage{stackengine}
\usepackage{fontawesome}
\usepackage{float}
\usepackage[hidelinks]{hyperref}
\usepackage{marvosym}

\def\techreportversion{}
\newcommand{\moreless}[2]{\ifdefined\techreportversion#1\else#2\fi}
\newcommand{\tighten}{\moreless{}{\looseness=-1}}
\newcommand{\techrep}[1]{\moreless{#1}{}}

\newcommand{\smartsec}[1]{\par\medskip\noindent\textbf{#1}\,}

\AtEndEnvironment{example}{\phantom{}\qed}
\AtEndEnvironment{proof}{\phantom{}\qed}

\usepackage[numbers,sort,sectionbib]{natbib} 
\makeatletter
\def\NAT@def@citea{\def\@citea{\NAT@separator}}
\makeatother

\definecolor{colorBoxBg}{RGB}{242, 248, 248}
\definecolor{colorBoxFg}{RGB}{204, 227, 222}
\definecolor{ProofColor}{RGB}{137, 83, 141}
\definecolor{colorRelevant}{RGB}{40,40,220}
\definecolor{colorFault}{RGB}{200,20,20}
\definecolor{colorWitnessA}{RGB}{20, 100, 180}
\definecolor{colorWitnessB}{RGB}{110, 160, 10}

\usepackage{cleveref}
\crefformat{section}{\S#2#1#3}
\crefmultiformat{section}{\S#2#1#3}{ and~\S#2#1#3}{, \S#2#1#3}{, and~\S#2#1#3}
\crefrangeformat{section}{\S#3#1#4 to~\S#5#2#6}

\usepackage{mathpartir}

\usepackage{listings}
\usepackage{relsize}
\lstdefinestyle{inline}{
  basicstyle=\relscale{.95}\ttfamily,
  keywords={},
}
\newcommand{\code}[2][]{\lstinline[style=inline,#1]!#2!}
\lstdefinestyle{mycode}{
  morekeywords={},
  deletekeywords={},
  lineskip=.1em,
  escapeinside={@}{@},
  numberstyle=\scriptsize,
  basicstyle=\scriptsize\ttfamily,
  columns=flexible,
  morecomment=[s][\color{green!60!black}]{/*}{*/},
  morecomment=[l][\color{green!50!black}]{//},
  moredelim=**[is][\color{purple}]{|<}{>|},
  mathescape=true,
  tabsize=2,
  firstnumber=last,
  numbersep=7pt,
  keepspaces=true,
  aboveskip=0pt,
  belowskip=0pt,
}
\makeatletter
\usepackage{tcolorbox}
\tcbset{
  traceStyle/.style={
    fonttitle=\footnotesize\ttfamily,
    coltitle=black,
    colback=colorBoxBg,
    colframe=colorBoxFg,
    size=fbox,
    arc=2mm,
    title={#1},
    toptitle=.25mm,
    bottomtitle=.25mm,
  }
}
\newtcolorbox{tracebox}[2][]{traceStyle={#2},#1}
\makeatother

\usepackage{tikz}
\usetikzlibrary{automata,positioning,arrows,calc,tikzmark}

\definecolor{colorMyGreen}{RGB}{80,170,0}
\definecolor{colorMyRed}{RGB}{180,20,20}
\definecolor{colorMyBlue}{RGB}{40,40,220}
\definecolor{colorMyPink}{RGB}{220,40,220}

\usepackage{stackengine}
\stackMath
\newcommand\xxrightarrow[1]{\raisebox{-.85pt}{\ensuremath{\smash{\mathrel{%
  \setbox2=\hbox{\stackon{\scriptstyle#1}{\scriptstyle#1}}%
  \stackon[-3.0pt]{%
    \xrightarrow{\makebox[\dimexpr\wd2\relax]{}}%
  }{%
   \scriptstyle#1\,%
  }%
}}}}}

\usepackage{xparse}
\makeatletter
\NewDocumentCommand{\tagx}{om}{%
  \IfNoValueTF{#1}
   {
    \refstepcounter{equation}(\theequation)\label{#2}%
   }
   {
    (#1)\def\@currentlabel{#1}\label{#2}%
   }%
}
\makeatother

\usepackage{pgfplots}
\DeclareUnicodeCharacter{2212}{−}
\usepgfplotslibrary{groupplots,dateplot}
\usetikzlibrary{patterns,shapes.arrows}
\pgfplotsset{compat=newest}
\usepackage{tikzscale}

\def\prallspacing{\mskip 2mu plus 2mu minus 3mu}
\newcommand{\prall}[1]{{\prallspacing{#1}\prallspacing}}

\newcommand{\true}{\mathit{true}}
\newcommand{\false}{\mathit{false}}
\renewcommand{\emptyset}{\varnothing}

\newcommand{\setcompact}[1]{\{#1\}}
\newcommand{\set}[1]{\{\,#1\,\}}

\newcommand{\setcond}[2]{\set{#1\;\mid\;#2}}
\newcommand{\cardof}[1]{|#1|}

\newcommand{\nat}{\mathbb{N}}
\newcommand{\ZZ}{\mathbb{Z}}
\newcommand{\RR}{\mathbb{R}}
\newcommand{\RRplus}{\RR_{\geq 0}}

\renewcommand{\to}{\rightarrow}
\newcommand{\defeq}{\triangleq}
\newcommand{\defiff}{{:}\!{\iff}}

\newcommand{\project}[2]{#1|_{#2}}

\newcommand{\vecof}[1]{\overline{#1}}

\newcommand{\nmodels}{\not\models}

\newcommand{\markBox}[2][2.5mm]{\parbox{#1}{#2~}}
\newcommand{\markR}{{\scalebox{.8}{\textcolor{colorRelevant}{\faCircle}}}\xspace}
\newcommand{\markF}{{\scalebox{.8}{\textcolor{colorFault}{\faBug}}}\xspace}
\newcommand\encircle[1]{\protect\tikz[baseline=(char.base)]{\protect\node[shape=circle,draw,inner sep=1pt,minimum height=3.5mm,minimum width=3.5mm] (char) {\texttt{#1}};}}
\newcommand{\classA}{\encircle{A}}
\newcommand{\classB}{\encircle{\kern+.5ptB}}

\newcommand{\anaut}{\mathcal{A}}
\newcommand{\setstates}{\mathit{Q}}
\newcommand{\astate}{\mathit{p}}
\newcommand{\astatep}{\mathit{q}}

\newcommand{\setinitial}{\mathit{S}}

\newcommand{\setvars}{\mathit{V}}
\newcommand{\avar}{\mathit{v}}

\newcommand{\setclocks}{\mathit{C}}
\newcommand{\aclock}{\mathit{c}}
\newcommand{\aclockp}{\mathit{c'}}
\newcommand{\setevents}{\mathit{E}}
\newcommand{\anevent}{\mathit{e}}

\newcommand{\timeevent}{\Delta}
\newcommand{\trans}[1][]{\rightarrow_{#1}}

\newcommand{\transofNEW}[4][]{\xxrightarrow{#2,\,#3,\,#4}_{#1}}
\newcommand{\stepofNEW}[2][]{\xxrightarrow{\,#2\,}_{#1}}
\newcommand{\aguard}{\mathit{g}}

\newcommand{\aconf}{\mathit{cf}}
\newcommand{\aconfp}{\mathit{cf'}}
\newcommand{\aconfpp}{\mathit{cf''}}
\newcommand{\aconfppp}{\mathit{cf'''}}
\newcommand{\amap}{\varphi}
\newcommand{\amapp}{\varphi'}
\newcommand{\amappp}{\varphi''}
\newcommand{\aword}{\mathit{w}}
\newcommand{\awordp}{\mathit{w'}}
\newcommand{\awordpp}{\mathit{w''}}
\newcommand{\asymbol}{\mathit{s}}

\newcommand{\awitness}{\widehat{w}}
\newcommand{\atime}{\mathit{t}}

\newcommand{\semof}[1]{\llbracket#1\rrbracket}
\newcommand{\ssemof}[3]{\llbracket#1\vert#2\rrbracket^\sharp_{#3}}
\newcommand{\smesof}[3]{\overline{\llbracket#1\vert#2\rrbracket}^\sharp_{#3}}

\newcommand{\anup}{\mathit{up}}

\newcommand{\noup}{\emptyset}

\newcommand{\langof}[1]{\mathcal{L}(#1)}

\newcommand{\aprop}{\mathit{P}}
\newcommand{\apropp}{\mathit{R}}
\newcommand{\aproppp}{\mathit{S}}
\newcommand{\sinit}{\mathit{Init}}
\newcommand{\aform}{\mathit{F}}
\newcommand{\aformp}{\mathit{G}}
\newcommand{\aformpp}{\mathit{H}}
\newcommand{\hoare}[3]{\set{#1}\,#2\,\set{#3}}
\newcommand{\wpre}[1][]{\mathit{wp}_{#1}}
\newcommand{\wpof}[3][]{\wpre[#1](#2,\,#3)}
\newcommand{\post}[1]{\mathit{post}^\sharp_{#1}}
\newcommand{\postof}[2]{\post{#1}(#2)}
\newcommand{\pre}[1]{\mathit{pre}^\sharp_{#1}}
\newcommand{\preof}[2]{\pre{#1}(#2)}
\newcommand{\spost}[1][]{\mathit{sp}_{#1}}
\newcommand{\spof}[3][]{\spost[#1](#2,\,#3)}
\newcommand{\asconf}{\aconf_{\!\sharp}}

\newcommand{\tequiv}{\equiv}
\newcommand{\widening}{\nabla}
\newcommand{\narrowing}{\overline{\nabla}}

\newcommand{\prel}[1][]{\sqsubseteq_{#1}}

\newcommand{\erel}[1][]{\sim_{#1}}
\newcommand{\anexp}{\alpha}
\newcommand{\anexpp}{\beta}
\newcommand{\anout}{\sigma}
\newcommand{\awitset}{\mathit{W}}
\newcommand{\atimevar}{\mathit{u}}

\newcommand{\tconsof}[1]{\mathit{acc}(#1)}
\newcommand{\issat}{\mathrm{SAT}}
\newcommand{\issatof}[1]{\issat(#1)}

\newcommand{\meet}{\sqcap}
\newcommand{\bigmeet}{\bigsqcap}
\newcommand{\aninterpolant}{\mathit{I}}
\newcommand{\interpolantof}[2]{\mathit{I}(#1,#2)}

\newcommand{\milis}{\mathrm{\textit{ms}}}

\newcommand{\exAutOp}{\ensuremath{\anaut_{\setevents}}}
\newcommand{\exAutDog}{\ensuremath{\anaut_{\timeevent}}}

\newcommand{\exMode}{\mathit{ctx}}
\newcommand{\exVal}{\text{\code{<val>}}}
\newcommand{\exData}{\text{\code{<data>}}}
\newcommand{\exClock}{\mathit{clk}}

%% file: content/abstract.tex

\begin{abstract}
Intensive testing using model-based approaches is the standard way of demonstrating the correctness of automotive software. Unfortunately, state-of-the-art techniques leave a crucial and labor intensive task to the test engineer: identifying bugs in failing tests. Our contribution is a model-based classification algorithm for failing tests that assists the engineer when identifying bugs. It consists of three steps. (i) Fault localization replays the test on the model to identify the moment when the two diverge. (ii) Fault explanation then computes the reason for the divergence. The reason is a subset of messages from the test that is sufficient for divergence. (iii) Fault classification groups together tests that fail for similar reasons. Our approach relies on machinery from formal methods: (i) symbolic execution, (ii) Hoare logic and a new relationship between the intermediary assertions constructed for a test, and (iii) a new relationship among Hoare proofs. A crucial aspect in automotive software are timing requirements, for which we develop appropriate Hoare logic theory. We also briefly report on our prototype implementation for the CAN bus \emph{Unified Diagnostic Services} in an industrial project.

\keywords{Fault Explanation \and Fault Classification \and Hoare Proofs.}

\par\addvspace\baselineskip\noindent
\textbf{Conference Version:}\enspace\ignorespaces
Becker, M., Meyer, R., Runge, T., Schaefer, I., van der Wall, S., Wolff, S.: \emph{Model-Based Fault Classification for Automotive Software}.
In: Sergey, I. (eds) Programming Languages and Systems. APLAS 2022. Lecture Notes in Computer Science, vol 13658. Springer, Cham. \url{https://doi.org/10.1007/978-3-031-21037-2_6}
\end{abstract}

%% file: content/base.tex

\input{content/intro}
\input{content/model}
\input{content/localization}
\input{content/explanation}
\input{content/classification}
\input{content/hoare}

\input{content/eval}

\input{content/related}

%% file: content/intro.tex

\section{Introduction}
\label{sec:intro}

\newcommand{\aquote}[1]{\emph{``#1''}}

Intensive testing is the de-facto standard way of demonstrating the correctness of automotive software, and the more tests the higher the confidence we have in a system~\cite{ASEBook2016}.
Model-based approaches have been instrumental in pushing the number of tests that can be evaluated, by increasing the degree of automation for the testing process.
Indeed, all of the following steps are fully automated today: determining the test cases including the expected outcome, running them on the system, and comparing the outcome to the expectation~\cite{MBTSurvey2012}.
Yet, there is a manual processing step left that, so far, has resisted automation.
If the outcome of the test and the expectation do not match, the bug has to be identified.
This is the moment the test engineer comes into play, and also the moment when automation strikes back.
The bug will not only show up in one, but rather in a large number of test cases, and the engineer has to go through all of them to make sure not to miss a mistake.
This is the problem we address: assist the test engineer when searching for bugs among a large number of failing tests.

Though our ideas may apply more broadly, we develop them in the context of hardware-in-the-loop testing for embedded controllers (ECUs) in the automotive industry~\cite{MBTAutomotive08}.
The final ECU with its software is given to the test engineer as a black box.
During testing, the ECU interacts with a (partly simulated) physical environment.
This interaction is driven by a test suite derived from a test model.
There are several characteristics that make hardware-in-the-loop testing substantially different from the earlier steps in the continuous integration and testing process (model/software/processor-in-the-loop testing).
The first is the importance of timing requirements~\cite{AD94}.
Second, the ECU with its software is a black-box.
Indeed, in our setting it is provided by a supplier and the testing unit does not have access to the development model.
Third, there is a test model capturing the product requirements document (PRD).
It is a complex artifact that specifies the intended system behavior at a fine level of detail, including logical states, transitions, timing requirements, and message payloads.
Indeed, \aquote{testing automotive systems often requires test scenarios with a very precise sequence of time-sensitive actions}~\cite{MBTAutomotive08}.
As is good practice~\cite{MBTAutomotive08,MBTSurvey2012,ASEBook2016}, the test model is different from the development model (it is even developed by a different company).
Lastly, there are hundreds to thousands of tests, which is not surprising as it is known that real-time requirements \aquote{are notoriously hard to test}~\cite{MBTSurvey2012}.

\begin{example}
	\label{ex:tests}
	\input{content/examples/traces}

	\Cref{fig:example-tests} illustrates the task at hand (ignore the \markR marks for now).
	The figure shows three traces derived from the \emph{Unified Diagnostic Services}~\cite{iso-uds}.
	A trace is a recording of the requests and responses resulting from executing a test case (pre-defined request sequence) on the ECU under test.
	Each line of the trace contains one message, carrying:
	\begin{inparaenum}
		\item a~time stamp indicating the time since the last message resp. the start,
		\item the type of message, \code{req} for requests and \code{res} for responses,
		\item an ECU identifier, the recipient for requests and the sender for responses,
		\item the name of an operation, e.g., \code{set}, and
		\item optional payload.
	\end{inparaenum}

	In the first trace, the ECU with identifier \code{CTR} is requested to perform the \code{set} operation with value \code{5}.
	The ECU acknowledges that the operation was executed successfully, repeating value \code{5}.
	Subsequently, \code{CTR} receives a \code{get} request to which it responds with (returns) value \code{0}.
	The second trace additionally requests a \code{log} operation between \code{set} and \code{get}.
	In the third trace, \code{get} returns \code{5} instead of \code{0}.

	The \code{get} responses in all traces are marked with \markF because they are faulty.
	Our example PRD requires \code{get} to return the value of the latest \code{set}, unless more than $50ms$ have passed since the latest (response~to) \code{set}, in which case \code{0} has to be returned.
	Assume the PRD does not specify any influence of \code{log} on \code{set}/\code{get}, and vice versa.
	The first two traces expose the same fault, indicated by \classA: the \code{set} appears to have been ignored.
	The last trace exposes a different fault, indicated by \classB: \code{CTR} appears to have ignored that $50ms$ have passed.
\end{example}

\tighten
Our contribution is an algorithm that classifies failing test cases according to their causes.
The algorithm expects as input the same information that is available to the test engineer:
the test model and the traces of the failing tests.
It consists of three steps: fault localization, fault explanation, and fault classification.
The fault localization can be understood as replaying a trace on the model to identify the moment when the two diverge.
In \Cref{ex:tests}, this yields the \markF marks.
The fault explanation then computes the reason for the divergence.
The reason can be understood as a small set of messages in the trace that is sufficient for the divergence.
In the example, this set is marked with \markR.
Even when removing the remaining messages, we would still have a bug.
The fault classification groups together traces that are faulty for similar reasons.
In the example, labels~$\classA$ and~$\classB$.

Our approach relies on machinery from formal methods, following the slogan in~\cite{MBTAutomotive08}: \aquote{more formal semantics are needed for test automation}.
Behind the fault localization is a symbolic execution~\cite{King76,Dart05}.
The challenge here is to summarize loops in which time passes but no visible events are issued.
We solve the problem with a widening approach from abstract interpretation~\cite{DBLP:conf/popl/CousotC77}.
Our fault explanation~\cite{DBLP:conf/sigsoft/Zeller02,DBLP:conf/spin/GroceV03,DBLP:conf/popl/BallNR03,DBLP:conf/kbse/RenierisR03,DBLP:conf/tacas/Groce04,Guo06,DBLP:conf/pldi/JoseM11} is based on Hoare logic~\cite{DBLP:conf/vmcai/ChristESW13,DBLP:conf/icse/Schwartz-Narbonne15}.
The challenge is to identify messages as irrelevant (for making the test fail), if they only let time pass but their effect is dominated by earlier parts of the test.
We achieve this using a new relationship between the assertions in the Hoare proof that is constructed for the test at hand.
The fault classification~\cite{Clustering09,RegressionSurvey2012} equates Hoare proofs~\cite{DBLP:conf/tap/PodelskiSW16}.
The challenge is again related to timing: the precise moments in which messages arrive will be different from test to test.
We propose a notion of proof template that allows us to equate Hoare proofs only based on timing constraints satisfied by the underlying tests.
The precise timing does not matter.

\tighten
\moreless{We implemented the classification in a project with the automotive industry. Our implementation targets the CAN bus \emph{Unified Diagnostic Services}.}{We implemented the classification in a project with the automotive industry, targeting the CAN bus \emph{Unified Diagnostic Services}.}
The test model has all the features mentioned above: real time, messages, and numerical payloads.
It is derived from a PRD with 350 pages of natural language and has 12k states and 70k transitions.
Our approach is practical: in 24 minutes we process test suites of up to 1000 tests with an average of 40 and outliers of up to 2500 messages in length.

\tighten
One may wonder why we classify tests at all.
Since they are derived from a test model, why not group them by the functionality they test or coverage they achieve?
The point is that functionality and coverage are only means of exposing faults~\cite{RegressionSurvey2012}.
The faults are what matters for the test engineer, and the same fault will show up in tests for different functions.
Our experiments confirm this: we discover previously undetected faults in tests that targeted functions different from the failing one.
We are particularly successful with faults involving timing, which are largely function independent and therefore admit a high degree of non-determinism.
Taking a step back, tests are designed by functionality or coverage, because it is hard to anticipate or even formulate possible faults in advance~\cite{Clustering09,RegressionSurvey2012,MBTSurvey2012,DBLP:journals/tse/WongGLAW16}.
Our explanation step makes the notion of a fault precise, and allows us to obtain the classification that the engineer needs for writing a test report.

Another question is whether we approach the problem from the wrong side.
There is a large body of work on test suite minimization~\cite{RegressionSurvey2012,TestingMLSurvey2022}.
So why classify tests a posteriori when we could have executed fewer tests in the first place?
The answer is that test suite minimization techniques are known to reduce the fault detection effectiveness, as demonstrated in the famous WHLM~\cite{WHLM}, WHMP~\cite{WHMP}, and Siemens~\cite{Siemens} studies.
This is unacceptable in the automotive sector.

\moreless{This technical report originally appeared as~\cite{published}.}{A companion technical report containing missing details is available as \cite{techreport}.}

%% file: content/examples/traces.tex

\begin{figure}[tb]
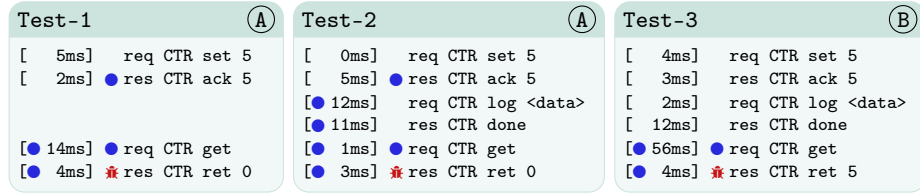

	\centering
	\begin{minipage}{.30\linewidth}
		\begin{tracebox}{Test-1\hfill\smash{\classA}}
\begin{lstlisting}
[@\markBox{      }@ 5ms] @\markBox{      }@req CTR set 5
[@\markBox{      }@ 2ms] @\markBox{\markR}@res CTR ack 5

[@\markBox{\markR}@14ms] @\markBox{\markR}@req CTR get
[@\markBox{\markR}@ 4ms] @\markBox{\markF}@res CTR ret 0
\end{lstlisting}
		\end{tracebox}
	\end{minipage}
	\hfill
	\begin{minipage}{.34\linewidth}
		\begin{tracebox}{Test-2\hfill\smash{\classA}}
\begin{lstlisting}
[@\markBox{      }@ 0ms] @\markBox{      }@req CTR set 5
[@\markBox{      }@ 5ms] @\markBox{\markR}@res CTR ack 5
[@\markBox{\markR}@12ms] @\markBox{      }@req CTR log <data>
[@\markBox{\markR}@11ms] @\markBox{      }@res CTR done
[@\markBox{\markR}@ 1ms] @\markBox{\markR}@req CTR get
[@\markBox{\markR}@ 3ms] @\markBox{\markF}@res CTR ret 0
\end{lstlisting}
		\end{tracebox}
	\end{minipage}
	\hfill
	\begin{minipage}{.34\linewidth}
		\begin{tracebox}{Test-3\hfill\smash{\classB}}
\begin{lstlisting}
[@\markBox{      }@ 4ms] @\markBox{      }@req CTR set 5
[@\markBox{      }@ 3ms] @\markBox{      }@res CTR ack 5
[@\markBox{      }@ 2ms] @\markBox{      }@req CTR log <data>
[@\markBox{      }@12ms] @\markBox{      }@res CTR done
[@\markBox{\markR}@56ms] @\markBox{\markR}@req CTR get
[@\markBox{\markR}@ 4ms] @\markBox{\markF}@res CTR ret 5
\end{lstlisting}
		\end{tracebox}
	\end{minipage}
	\vspace{-5pt}
	\caption{%
		Traces of an ECU \code{CTR} with operations \code{set}, \code{get}, and \code{log}.
		Faults are marked with \markF, relevant events with \markR.
		Labels \scalebox{.9}{$\classA$} and \scalebox{.9}{$\classB$} indicate distinct causes for the faults.
		\label{fig:example-tests}%
	}
	\vspace{-2.5mm}
\end{figure}

%% file: content/model.tex

\section{Formal Model}
\label{sec:model}

We introduce a class of automata enriched by memory and clocks to model PRDs.
A so-called \emph{PRD automaton} is a tuple $\anaut=(\setstates, \trans, \setinitial, \setevents, \setvars, \setclocks)$ with a finite set of states $\setstates$, a finite transition relation $\trans$ among states, initial states $\setinitial\subseteq\setstates$, a finite set of events $\setevents$, a finite set of memory variables $\setvars$, and a finite set of clocks $\setclocks$.
Variables and clocks are disjoint, $\setvars\cap\setclocks=\emptyset$.
Transitions take the form $\astate\transofNEW{\anevent}{\aguard}{\anup}\astatep$ with states $\astate,\astatep\in\setstates$, event $\anevent\in\setevents$, guard $\aguard$, and update $\anup$.
Additionally, there are transitions $\astate\transofNEW{\timeevent}{\aguard}{\anup}\astatep$ that react on time progression, denoted by the special symbol $\timeevent\notin\setevents$.
\emph{Guards} are Boolean formulas over (in)equalities of memory variables, clocks, and constants.
We assume a strict typing and forbid (in)equalities among memory variables and clocks.
\emph{Updates} are partial functions that may give new values to variables $\avar$, $\anup(\avar)\in\ZZ$, or reset clocks $\aclock$, $\anup(\aclock)=0$.
Lifting variable updates from values to terms (over variables) is straightforward.

The runtime behavior of PRD automata is defined in terms of labeled transitions between configurations.
A \emph{configuration} of $\anaut$ is a tuple $\aconf=(\astate,\amap)$ consisting of a state $\astate\in\setstates$ and a total valuation $\amap:{\setvars{\,\to\;}\ZZ}~\,{\cup}\;{\setclocks{\;\to\;}\RRplus}$ of variables and clocks.
The configuration is initial if $\astate\in\setinitial$ is initial (no constraints on $\amap$).

Valuations $\amap$ are affected by the progression of time $\atime$ and updates $\anup$.
Progressing $\amap$ by $\atime$ yields a new valuation $\amap + \atime$, coinciding with $\amap$ on all variables $\avar$ and advancing all clocks $\aclock$ by $t$, ${(\amap + \atime) (\aclock) = \amap(\aclock) + \atime}$.
To apply $\anup$ to $\amap$, we introduce the \emph{transformer} $\semof{\anup}$.
It yields a new valuation $\semof{\anup}(\amap)=\amapp$ such that \[
	\newcommand{\tre}[3]{{#1}{\:\,{?}\:\,}{#2}{\:\,{:}\:\,}{#3}}
	\amapp(\avar) \;=\; \tre{ \anup(\avar)\neq\bot }{ \anup(\avar) }{ \amap(\avar) }
	\quad~\text{and}~\quad
	\amapp(\aclock) \;=\; \tre{ \anup(\aclock)\neq\bot }{ 0 }{ \amap(\aclock) }
	\ .
\]

PRD automata $\anaut$ process finite traces $\aword=\asymbol_1\dots\asymbol_n$ of events and time progressions, $\asymbol_i\in\setevents\cup\RRplus$.
Events are instantaneous and time progressions make explicit the passing of time.
A \emph{basic run} $(\astate_1,\amap_1)\stepofNEW{\asymbol_1}\cdots\stepofNEW{\asymbol_n}(\astate_{n+1},\amap_{n+1})$ of $\anaut$ on $\aword$ is a sequence of steps where $(\astate_1,\amap_1)$ is initial.
Steps $(\astate,\amap)\stepofNEW{\anevent}(\astatep,\amapp)$ for events $\anevent\in\setevents$ are due to transitions in $\anaut$, so they satisfy the following two conditions:
\begin{compactenum}[(i)]
	\item
		There is a transition $\astate\transofNEW{\anevent}{\aguard}{\anup}\astatep$ such that $\aguard$ is enabled.
		Enabledness means that $\amap$ is a model of $\aguard$, written $\amap\prall{\models}\aguard$.
	\item
		The valuation $\amapp$ is induced by the transformer for $\anup$, $\amapp=\semof{\anup}(\amap)$.
\end{compactenum}
Similarly, steps $(\astate,\amap)\stepofNEW{\atime}(\astatep,\amapp)$ taking time $\atime\in\RRplus$ require:
\begin{compactenum}[(i)]
	\item
		There is a $\timeevent$-transition $\astate\transofNEW{\timeevent}{\aguard}{\anup}\astatep$ enabled after waiting $\atime$ time, $\amap + \atime \models\aguard$.
	\item
		Valuation $\amapp$ is induced by clock progression plus $\anup$, $\amapp=\semof{\anup}(\amap + \atime)$.
\end{compactenum}
Finally, there are stuttering steps $(\astate, \amap) \stepofNEW{0} (\astate, \amap)$ which have no requirements.

Next, we lift basic runs to allow for multiple $\timeevent$-transitions during a single time progression $\atime$ in $\aword$.
This is needed to support complex behavior while waiting, as seen in \Cref{ex:tests}.
We rewrite $\aword$ by splitting and merging time progressions.
More precisely, we rewrite $\aword$ into $\awordp$ along these equivalences:
\begin{align*}
	\aword_1.\aword_2\tequiv\aword_1.0.\aword_2
	\qquad\text{and}\qquad
	\aword_1.\atime.\aword_2\tequiv\aword_1.\atime_1.\atime_2.\aword_2 \text{~~if~~} \atime=\atime_1+\atime_2
	\ .
	\tag{TEQ}
	\label{eq:trace-equalities}
\end{align*}
Then, we say that $\anaut$ \emph{has a run on} $\aword$ if there is $\awordp$ with $\awordp\tequiv\aword$ so that $\anaut$ has a basic run on $\awordp$.
The specification $\langof{\anaut}$ induced by $\anaut$ is the set of all traces $\aword$ on which $\anaut$ has a run.
Readers familiar with hybrid systems will observe that our rewriting produces finite decompositions only, thus excludes zeno behavior~\cite{DBLP:conf/rex/AbadiL91}.

To simplify the exposition, we hereafter implicitly assume that traces $\aword$ are normalized in the sense that every event is preceded and succeeded by exactly one time progression.
This normalization is justified by the \eqref{eq:trace-equalities}~equivalences.

\tighten
In practice, models have many transitions between two states in order to capture state changes that ignore parts of the event or accept a large number of possible values.
To avoid PRD automata growing unnecessarily large, we use regular expressions instead of single events as transition labels.
The automaton model presented so far naturally extends to such a lift.
Our implementation integrates this optimization, see~\Cref{sec:iav}.
For simplicity, we stick to vanilla automata hereafter.

\vspace{-1mm}
\begin{example}
	\label{ex:automata}
	\tighten
	The automata $\exAutOp,\exAutDog$ from \Cref{fig:example-aut} specify \code{CTR} from \Cref{ex:tests}.
	Automaton $\exAutOp$ addresses \code{get}, \code{log}, and \code{set}.
	The \code{set} request takes an arbitrary value $\exVal$ as a parameter.
	As discussed above, we use $\exVal$ as shorthand which can be translated on-the-fly into vanilla automata.
	The \code{set} request is always enabled and does not lead to updates.
	It may be followed by an \code{ack}, indicating success, or a \code{fail} response.
	If successful, variable $\exMode$ is updated to $\exVal$.
	The reset of $\exMode$ after $50ms$ is implemented by $\exAutDog$.
	Operations \code{get} and \code{log} are~similar.
	
	\input{content/examples/automata}

	Automaton $\exAutOp$ does not specify any timing behavior, all its states have an always-enabled $\timeevent$-self-loop without updates.
	The timing behavior is specified by automaton $\exAutDog$.
	It uses \code{ack} responses as a trigger to reset the timer $\exClock$ and then waits until $\exClock$ holds a value of at least $50$.
	Once the threshold is reached, the $\timeevent$-transition from $\astate_4$ to $\astate_5$ setting $\exMode$ to $0$ becomes enabled.
	Here, $\exAutDog$ allows for slack: the reset must happen within $5ms$ once $50ms$ have passed.
	Within these $5ms$, $\exAutDog$ may choose to cycle in $\astate_4$ without resetting or move to $\astate_5$ while resetting $\exMode$.
	In practice, this kind of slack is common to account for the inability of hardware to execute after exactly $50ms$, as a guard like $\exClock\leq50$ would require.

	The overall specification of our example is the composition $\exAutOp\times\exAutDog$.
	The cross-product is standard: a step can be taken only if both $\exAutOp$ and $\exAutDog$ can take the step.
	We do not go into the details of operations over automata.
\end{example}

%% file: content/examples/automata.tex

\begin{figure}[tb]
	\centering
	\newcommand{\myVar}[1]{\mathit{#1}}
	\newcommand{\myRef}[1]{\text{\code{#1}}}
	\newcommand{\myEvent}[2][]{\code{#2}#1}
	\newcommand{\myTimeEvent}{$\timeevent$}
	\newcommand{\myStuff}[2]{$#1$, #2}
	\newcommand{\myAction}[1]{$\setcompact{#1}$}
	\newcommand{\myNoAction}{$\noup$}
	\newcommand{\myMode}{\exMode}
	\newcommand{\myVal}{\exVal}
	\newcommand{\myData}{\exData}
	\newcommand{\myClock}{\exClock}
	\newcommand{\addSelfLoop}[2][]{
		\draw[->] (#2) edge[loop below,min distance=5.85mm,out=235,in=305]
			node[above,yshift=-0.7pt] {\myTimeEvent}
			node[below] {#1}
			(#2);
	}
	\newcommand{\addSelfLoopTop}[2]{
		\draw[->] (#1) edge[loop above,min distance=6.75mm,out=125,in=55]
			node[below] {#2}
			(#1);
	}
	\tikzstyle{autstyle}=[>=stealth,thick,x=3.85mm,y=-7.0mm,every state/.style={fill=gray!10,minimum size=5mm},initial text={}]
	\begin{tikzpicture}[autstyle]
		\scriptsize


		\node[state,initial,label={[anchor=south east,xshift=-2mm,yshift=-2mm]\exAutOp}] (init) at (0,0) { $\astate_0$ };
		\node[state] (set) at (11.5,2) { $\astate_1$ };
		\node[state] (log) at (15.5,0) { $\astate_3$ };
		\node[state] (get) at (13.5,1) { $\astate_2$ };
		\node[state,dashed,fill=white] (fin) at (28.5,0) { $\astate_0$ };
		
		\draw[->] (init) -- (log);
		\draw[->] (1.25,0) |- (get);
		\draw[->] (1.25,1) |- (set);
		\draw[->] (log) -- (fin);
		\draw (13.2,2) -- (13.2,2.9) -| (28.5,2);
		\draw[->] (get) -| (fin);
		\draw (set) -| (28.5,1);
		\addSelfLoop{init}
		\addSelfLoop{set}
		\addSelfLoop{log}
		\addSelfLoop{get}

		\path (0,2) --
			node[above] {\myEvent[ \myVal]{req CTR set}, \myStuff{\true}{\myNoAction}}
			(12,2);
		\path (15,2) --
			node[above] {\myEvent[ \myVal]{res CTR ack}, \myStuff{\true}{\myAction{\myMode\mapsto\myVal}}}
			(28,2);
		\path (15,2.9) --
			node[above] {\myEvent[ \myVal]{res CTR fail}, \myStuff{\true}{\myNoAction}}
			(28,2.9);

		\path (0,0) --
			node[above] {\myEvent[ \myData]{req CTR log}, \myStuff{\true}{\myNoAction}}
			(12,0);
		\path (15,0) --
			node[above] {\myEvent{res CTR done}, \myStuff{\true}{\myNoAction}}
			(28,0);

		\path (0,1) --
			node[above] {\myEvent{req CTR get}, \myStuff{\true}{\myNoAction}}
			(12,1);
		\path (15,1) --
			node[above] {\phantom{iiiii}\myEvent[ \myVal]{res CTR ret}, \myStuff{\myMode=\myVal}{\myNoAction}}
		 	(28,1);

	\end{tikzpicture}\\[-4mm]
	\begin{tikzpicture}[autstyle]
		\scriptsize
		\node[state,initial,label={[anchor=south east,xshift=-2mm,yshift=-2mm]\exAutDog}] (dInit) at (0,4) { $\astate_4$ };
		\node[state] (dReset) at (16.2,4) { $\astate_5$ };
		\node[state,dashed,fill=white] (dFin) at (24,4) { $\astate_4$ };

		\addSelfLoop[\myStuff{\myClock<55}{\myNoAction}]{dInit}
		\addSelfLoop{dReset}
		\addSelfLoopTop{dInit}{$\setevents\smash{'}$}
		\draw[->] (dInit) -- (2.3,4) --
			node[above] {\myTimeEvent, \myStuff{50\leq\myClock<55}{\myAction{\myMode\mapsto0}}}	
			(dReset);
		\draw[->] (2.3,4) -- (2.3,5) --
			node[above] {\myEvent[ \myVal]{res CTR ack}, \myStuff{\true}{\myAction{\myClock\mapsto0}}}
			(15.6,5) -| (dFin);
		\draw[->] (dReset) --
		 	node[above]{\myTimeEvent, \myStuff{\true}{\myNoAction}}
			(dFin);
	\end{tikzpicture}\\[-3mm]
	\caption{%
		Model $\exAutOp\times\exAutDog$ for the ECU \code{CTR} from \Cref{ex:tests}.
		Automaton $\exAutOp$ specifies operations \code{log}, \code{get}, and \code{set}.
		Automaton $\exAutDog$ specifies how variable $\exMode$ is reset.
		We omit the guards $\true$ and updates $\noup$ on $\timeevent$-loops.
		We use $\setevents'\defeq\setevents\setminus\setcompact{\text{\code{res CTR ack} \exVal}}$.%
		\label{fig:example-aut}%
	}
	\vspace{-1.5mm}
\end{figure}

%% file: content/localization.tex

\section{Fault Localization}
\label{sec:localization}

\tighten
We propose a method for localizing faults in traces $\aword$.
Intuitively, we do so by letting $\anaut$ run on $\aword$.
If for some prefix $\awordp\!.\asymbol$ of $\aword$ there is no step to continue the run, i.e., $\awordp\in\langof{\anaut}$ but $\awordp\!.\asymbol\notin\langof{\anaut}$, then $\asymbol$ is a fault and $\awordp\!.\asymbol$ is its \emph{witness}.
Witnesses play an integral role in our approach: a Hoare proof for a witness yields a formal reason for the fault.
In \Cref{sec:explanation}, we will refine this reason by extracting a concise explanation for the fault.
This explanation then allows us to classify faults in \Cref{sec:classification}.

Technically, identifying faults $\asymbol$ in $\aword$ is more involved.
Establishing $\awordp\in\langof{\anaut}$ requires us to find $\awordpp\tequiv\awordp$ and a basic run of $\anaut$ on $\awordpp$.
Establishing $\awordp\!.\asymbol\notin\langof{\anaut}$, however, requires us to show that there exists no basic run of $\anaut$ on $\awordp\!.\asymbol$ at all. 
It is not sufficient to show that the single basic run witnessing $\awordp\in\langof{\anaut}$ cannot be extended to $\awordp\!.\asymbol$.
We have to reason over all $\tilde \aword\tequiv\awordp\!.\asymbol$ and over all basic runs on them. 
To cope with this, we encode symbolically all such basic runs of $\anaut$ as a Hoare proof.
The Hoare proof can be thought of as a certificate for the fault.

Interestingly, our techniques for fault localization~(\Cref{sec:localization}), explanation~(\Cref{sec:explanation}), and classification~(\Cref{sec:classification}) do not rely on the exact form of Hoare proofs or how they are obtained---any valid proof will do.
Hence, we prefer to stay on the \emph{semantic level}.
We discuss how to efficiently generate the necessary proofs in \Cref{sec:hoare-theory}.
Note that the timing aspect of our model requires us to develop novel Hoare theory in \Cref{sec:hoare-theory}.

\smartsec{Symbolic Encoding}
We introduce a symbolic encoding to capture infinitely many configurations in a finite and concise manner.

A \emph{symbolic configuration} is a pair $\asconf=(\astate,\aform)$ where $\astate$ is a state and $\aform$ is a first-order formula.
We use $\aform$ to encode potentially infinitely many variable/clock valuations $\amap$.
We say $\aform$ denotes $\amap$ if $\amap$ is a model for $\aform$, written $\amap\models\aform$.

A \emph{condition} $\aprop$ is a finite set of symbolic configurations.
We write $(\astate,\amap)\models\aprop$ if there is $(\astate,\aform)\in\aprop$ with $\amap\models\aform$.
We also write $\aprop\prel\apropp$ if $\aconf\models\aprop$ implies $\aconf\models\apropp$ for all $\aconf$.
If $\aprop\prel\apropp$ and $\apropp\prel\aprop$, we simply write $\aprop=\apropp$.
The initial condition is $\sinit\defeq\setcond{(\astate,\true)}{\astate\in\setinitial}$ and the empty condition is $\false=\emptyset$.
For simplicity, we assume that conditions contain exactly one symbolic configuration per state, as justified by the next \namecref{thm:prop-rewriting}.
With that assumption, checking $\aprop\prel\apropp$ can be encoded as an SMT query and discharged by an off-the-shelf solver like \texttt{Z3}~\cite{DBLP:conf/tacas/MouraB08}.

\begin{lemma}
	\label{thm:prop-rewriting}
	$\aprop\cup\setcompact{(\astate,\aform),(\astate,\aformp)} = \aprop\cup\setcompact{(\astate,\aform\vee\aformp)}$
	and
	$\aprop\cup\setcompact{(\astate,\false)} = \aprop$.
\end{lemma}

Later, we will use conditions $\aprop$ below quantifiers $\exists\vecof{x}.\aprop$ and in the standard Boolean connectives $\aformp\oplus\aprop$ with formulas $\aformp$.
We lift those operations to conditions by pushing them into the symbolic configurations of $\aprop$ as follows:
\[
	~\exists\,\vecof{x}.\;\aprop \:\defeq\: \setcond{\! (\astate,\exists\,\vecof{x}.\;\aform) \!\!}{\!\! (\astate,\aform)\in\aprop \!}
	\quad\text{and}\quad
	\aformp\,{\oplus}\,\aprop \:\defeq\: \setcond{\! (\astate,\aformp\,{\oplus}\,\aform) \!\!}{\!\! (\astate,\aform)\in\aprop \!}
	\,.~
\]

\smartsec{Finding Faults}
We localize faults in traces $\aword=\asymbol_1\dots\asymbol_n$.
This means we check whether or not $\anaut$ has a run on $\aword$.
To do so, we rely on a Hoare proof for $\aword$ which takes the form \[
	\set{\aprop_0}\; \asymbol_1 \;\cdots \;\set{\aprop_{i-1}}\; \asymbol_i \;\set{\aprop_{i}}\; \cdots\; \asymbol_n \;\set{\aprop_n}
	\enspace,
\]
where every triple $\hoare{\aprop_i}{\asymbol_i}{\aprop_{i+1}}$ is a Hoare triple.
Intuitively, the Hoare triple means: every step for $\asymbol_i$ starting in a configuration from $\aprop_i$ leads to a configuration in $\aprop_{i+1}$.
Hoare triples are defined to be insensitive to trace equivalence: \[
	~\models\hoare{\aprop}{\asymbol}{\apropp}
	~~\defiff~~
	\forall\aconf,\aconfp,\awordp.~~
	\aconf\models\aprop
	\wedge
	\asymbol\equiv\awordp
	\wedge
	\aconf\stepofNEW{\awordp}\aconfp
	\implies
	\aconfp\models\apropp
	\ . ~
\]
If the condition is satisfied, we call the Hoare triple \emph{valid}.
For brevity, we write $\hoare{\aprop}{\awordp.\asymbol}{\aproppp}$ if there is $\apropp$ so that $\hoare{\aprop}{\awordp}{\apropp}$ and $\hoare{\apropp}{\asymbol}{\aproppp}$ are both valid.
Strengthening resp. weakening the precondition $\aprop$ resp. postcondition $\apropp$ preserves validity:
$\aprop'\prel\aprop$ and $\models\hoare{\aprop}{\asymbol}{\apropp}$ and $\apropp\prel\apropp'$ implies $\models\hoare{\aprop'}{\asymbol}{\apropp'}$.

Now, finding faults boils down to checking the validity of Hoare triples.
It is easy to see that $\anaut$ has no run on $\awordp\!.\asymbol$ if and only if $\models\hoare{\sinit}{\awordp\!.\asymbol}{\false}$.
\begin{lemma}
	\label{thm:hoare-characterizes-runs}
	If $\models\mkern-4mu\hoare{\mkern-2mu\sinit\mkern-2mu}{\mkern-.75mu\awordp\mkern-.75mu\mkern-1mu\,\set{\mkern-2mu\aprop\mkern-2mu}\,\mkern-.75mu\asymbol\mkern-.75mu}{\mkern-2mu\false\mkern-2mu}\mkern-2mu$ and $\aprop\prall{\neq}\false\mkern-.5mu$, then $\awordp\!.\asymbol$ witnesses fault~$\asymbol$.
\end{lemma}

\begin{example}
	\input{content/examples/proof}
	\Cref{fig:example-proofs} gives proofs that perform fault localization in \code{Test-1} and \code{Test-2} from \Cref{fig:example-tests}.
	The beginning of both traces is irrelevant for the fault, so $\true$ is used as precondition.
	Then, the conditions track the amount of time that passes in the form of an upper bound on clock $\exClock$.
	Since $\exClock$ stays below $50\milis$, variable $\exMode$ is never reset by $\exAutDog$.
	Hence, \code{get} must not return~$0$.
	But because \code{get} does return $0$ in the trace, we arrive at $\false$---the response is a fault.
\end{example}

The Hoare proof certifying witness $\awordp\!.\asymbol$ is input to the fault explanation and classification in the next sections.
As stated earlier, we defer the generation of Hoare proofs (by means of strongest postconditions and weakest preconditions) to \Cref{sec:hoare-theory}, as it is orthogonal to fault explanation and classification.

%% file: content/examples/proof.tex
{
\begingroup
\addtolength{\jot}{-0.2em}
\newcommand{\sconf}[3]{(\astate_{#1},\astate_{#2}:{#3})}
\newcommand{\sconfal}[3]{(\astate_{#1},\astate_{#2}&:{#3})}
\newcommand{\clk}{\exClock}
\newcommand{\ctx}{\exMode}
\renewcommand{\set}[1]{{\textcolor{ProofColor}{\{#1\}}}}
\begin{figure}[tb]
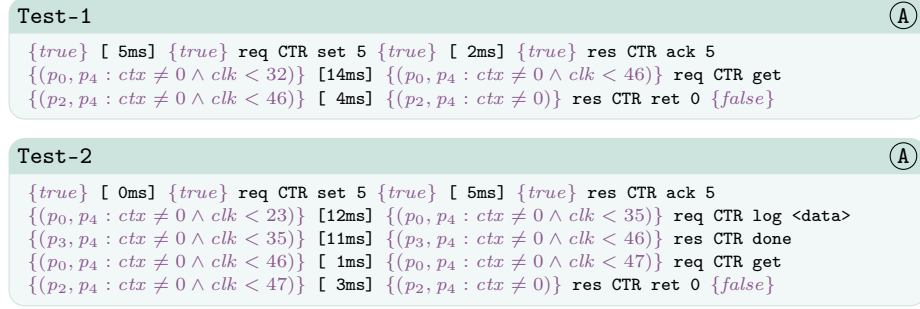

    \begin{tracebox}{Test-1\hfill\smash{\classA}}
\begin{lstlisting}[mathescape, escapeinside={(*}{*)}]
 $\set{\true}$ [ 5ms] $\set{\true}$ req CTR set 5 $\set{\true}$ [ 2ms] $\set{\true}$ res CTR ack 5
 $\set{\sconf{0}{4}{\ctx \neq 0 \land \clk < 32}}$ [14ms] $\set{\sconf{0}{4}{\ctx \neq 0 \land \clk < 46}}$ req CTR get
 $\set{\sconf{2}{4}{\ctx \neq 0 \land \clk < 46}}$ [ 4ms] $\set{\sconf{2}{4}{\ctx \neq 0}}$ res CTR ret 0 $\set{\false}$
\end{lstlisting}
    \end{tracebox}
    \begin{tracebox}{Test-2\hfill\smash{\classA}}
\begin{lstlisting}[mathescape, escapeinside={(*}{*)}]
 $\set{\true}$ [ 0ms] $\set{\true}$ req CTR set 5 $\set{\true}$ [ 5ms] $\set{\true}$ res CTR ack 5
 $\set{\sconf{0}{4}{\ctx \neq 0 \land \clk < 23}}$ [12ms] $\set{\sconf{0}{4}{\ctx \neq 0 \land \clk < 35}}$ req CTR log <data>
 $\set{\sconf{3}{4}{\ctx \neq 0 \land \clk < 35}}$ [11ms] $\set{\sconf{3}{4}{\ctx \neq 0 \land \clk < 46}}$ res CTR done
 $\set{\sconf{0}{4}{\ctx \neq 0 \land \clk < 46}}$ [ 1ms] $\set{\sconf{0}{4}{\ctx \neq 0 \land \clk < 47}}$ req CTR get
 $\set{\sconf{2}{4}{\ctx \neq 0 \land \clk < 47}}$ [ 3ms] $\set{\sconf{2}{4}{\ctx \neq 0}}$ res CTR ret 0 $\set{\false}$
\end{lstlisting}
    \end{tracebox}\vspace{-5pt}
    \caption{%
		Hoare proofs for \code{Test-1} and \code{Test-2}. 
        \label{fig:example-proofs}%
	}
\end{figure}
\endgroup
}%

%% file: content/explanation.tex

\section{Fault Explanation}
\label{sec:explanation}

We analyze the Hoare proof generated in \Cref{sec:localization} which certifies the fault in a witness.
Our goal is to extract the events that contribute to the fault and dispose of those that are irrelevant.
The result will be another valid Hoare proof that concisely explains the fault.
On the one hand, the explanation will help the test engineer understand the fault and ultimately prepare the test report alluded to in \Cref{sec:intro}.
On the other hand, explanations of distinct test cases may be similar in terms of our classification approach from \Cref{sec:classification} while the original test cases are not, thus improving the effectiveness of the classification.

To determine a concise explanation, assume the Hoare proof certifying the fault can be partitioned into \(\hoare{\sinit}{\aword_1}{\aprop}\,\aword_2\,\hoare{\apropp}{\aword_3}{\aprop_k}\).
If $\aprop$ denotes fewer configurations than $\apropp$, $\aprop\prel\apropp$, we say that $\aword_2$ is irrelevant (the events therein).
To see this, consider some configuration $\aconf\models\aprop$.
Executing $\aword_2$ from $\aconf$ leads to some $\aconfp\models\apropp$ which in turn leads to the fault by executing $\aword_3$.
However, $\aconf\models\apropp$ already holds.
So, we can just execute $\aword_3$ from $\aconf$ to exhibit the fault---$\aword_2$ is irrelevant indeed.

When timing plays a role in the fault, one might not be able to establish the simple inclusion $\aprop\prel\apropp$ because removing $\aword_2$ altogether also removes the time that passes in it.
However, it might be this passing of time, rather than the events, that leads to the fault.
Therefore, we also check if the events (and the events only) in $\aword_2$ are irrelevant.
This is the case if waiting has the same effect as performing full~$\aword_2$.
Technically, we check the validity of the triple $\hoare{\aprop}{\project{\aword_2}{\RRplus}}{\apropp}$.
The projection $\project{\aword_2}{\RRplus}$ removes all events $\setevents$ from $\aword_2$: $\project{\anevent}{\RRplus}=\epsilon$ and $\project{\atime}{\RRplus}=\atime$.
The validity of the triple captures our intuition: any configuration $\aconf\models\aprop$ can simply wait (taking $\timeevent$-transitions) for the same amount as $\aword_2$ and arrive in $\aconfp\models\apropp$ from which $\aword_3$ and the fault are executable---the removed events $\project{\aword_2}{\setevents}$ are irrelevant.

We apply the above reasoning---both $\aprop\prel\apropp$ as well as $\models\hoare{\aprop}{\project{\aword_2}{\RRplus}}{\apropp}$---to all partitionings of the given proof to identify the irrelevant sequences.
The remaining events and time progressions all contribute to the fault.
The result is the most concise explanation of the fault.

Unfortunately, our pruning rules are not confluent, meaning that different sequences of \emph{irrelevance checks} may lead to different explanations.
A witness may have more than one explanation if two irrelevant sequences partially overlap.
To see this, consider the following (special case) partitioning of the witness' proof 
{\setlength{\abovedisplayskip}{1.2em}\setlength{\belowdisplayskip}{1.2em}%
\[
	\newcommand{\myAnnot}[2]{\tikzmarknode{#1}{\set{\strut#2}}}
	\myAnnot{witI}{\sinit}\;
	\aword_1
	\;\myAnnot{witPa}{\aprop}\;
	\aword_2
	\;\myAnnot{witRa}{\apropp}\;
	\aword_3
	\;\myAnnot{witPb}{\aprop}\;
	\aword_4
	\;\myAnnot{witRb}{\apropp}\;
	\aword_5
	\;\myAnnot{witF}{\false}
	\ .
	\begin{tikzpicture}[overlay,remember picture,>=stealth,thick]
		\tikzset{my curve 7/.style args={#1of#2}{
			to path={.. controls ($(\tikztostart)!#2!(\tikztotarget)!#1!90:(\tikztotarget)$) 
			and ($(\tikztostart)!1-#2!(\tikztotarget)!#1!90:(\tikztotarget)$) 
			.. (\tikztotarget)\tikztonodes}},
			my curve 7/.default={7mm of 0.25}
		}
		\draw[->,color=colorWitnessA,opacity=.75] (witI.70) to [my curve 7 = 2.75mm of 0.12] (witPa.110);
		\draw[->,color=colorWitnessA,opacity=.75] (witPa.70) to [my curve 7 = 2.75mm of 0.12] (witPb.110);
		\draw[->,color=colorWitnessA,opacity=.75] (witPb.70) to [my curve 7 = 2.75mm of 0.12] (witRb.110);
		\draw[->,color=colorWitnessA,opacity=.75] (witRb.70) to [my curve 7 = 2.75mm of 0.12] (witF.110);
		\draw[->,color=colorWitnessB,opacity=.75] (witI.290) to [my curve 7 = -2.75mm of 0.12] (witPa.250);
		\draw[->,color=colorWitnessB,opacity=.75] (witPa.290) to [my curve 7 = -2.75mm of 0.12] (witRa.250);
		\draw[->,color=colorWitnessB,opacity=.75] (witRa.290) to [my curve 7 = -2.75mm of 0.12] (witRb.250);
		\draw[->,color=colorWitnessB,opacity=.75] (witRb.290) to [my curve 7 = -2.75mm of 0.12] (witF.250);
	\end{tikzpicture}
\]}%
Here, we deem irrelevant \textcolor{colorWitnessA}{$\aword_2.\aword_3$} and \textcolor{colorWitnessB}{$\aword_3.\aword_4$}.
However, we cannot remove $\aword_2.\aword_3.\aword_4$ entirely because the resulting proof might not be valid, which requires $\aprop\prel\apropp$.
Even removing the intersection $\aword_3$ of the irrelevant sequences may not produce a valid proof as $\apropp\prel\aprop$ might not hold either.
The same problems arise if only $\project{(\aword_2.\aword_3)}{\setevents}$ and/or $\project{(\aword_3.\aword_4)}{\setevents}$ is irrelevant.
We argue that this is desired: the witness is, in fact, a witness for two different faults, explained by \textcolor{colorWitnessA}{$\aword_1.\aword_4.\aword_5$} resp. \textcolor{colorWitnessB}{$\aword_1.\aword_2.\aword_5$}.
Overall, we compute all explanations in case there are overlapping irrelevant sequences.
While this gives exponentially many explanations in theory, we rarely find overlaps in practice.

\begin{example}
	\label{ex:explanation}
	We give the fault explanation for the proof of \code{Test-2} from \Cref{fig:example-proofs}.
	As expected, both events \code{req CTR log <data>} and \code{res CTR done} are irrelevant.
	The condition $\aprop = \set{(\astate_{0},\astate_{4}:{\exMode \neq 0 \land \exClock < 23})}$ before the \code{log} request reaches condition $\apropp = \set{(\astate_{0},\astate_{4}:{\exMode \neq 0 \land \exClock < 47})}$ after the \code{log} response.
	This remains true after removing both events.
	Indeed, $\hoare{\aprop}{\text{\code{[12ms][11ms][ 1ms]}}}{\apropp}$ is a valid Hoare triple and thus justifies removing the events.
\end{example}

%% file: content/classification.tex

\section{Fault Classification}
\label{sec:classification}

We propose a classification technique that groups together witnesses exhibiting the same or a similar fault.
Grouping together similar faults significantly reduces the workload of test engineers when preparing a test report for a large number of failing tests since only one (representative) test case per group needs to be inspected.
The input to our classification is a set~$\awitset$ of witness explanations as constructed in \Cref{sec:explanation}.
The result of the classification is a partitioning of~$\awitset$ into disjoint classes~$\awitset\prall{=}\awitset_1\uplus\cdots\uplus\awitset_m$.
The partitioning is obtained by factorizing~$\awitset$ along an equivalence~$\erel$ that relates witness explanations which have similar faults.
If~$\erel$ is effectively computable, so is the factorization.
We focus on~$\erel$.

Intuitively, two explanations are similar, and thus related by~$\erel$, if comprised of the same sequence of Hoare triples, that is, the same sequence of events and intermediary assertions.
This strict equality, however, does not work well when timing is involved.
Repeatedly executing the same sequence of events is expected to observe a difference in timing due to fluctuations in the underlying hardware.
Moreover, explanations have already been stripped by irrelevant sequences the events and duration of which might differ across explanations.

To make up for these discrepancies, we relate explanations that are equal up to \emph{similar clocks}.
Consider an (in)equality~$\aform$ over clocks~$\setclocks$.
We can think of~$\aform$, more concretely its solutions, as a polytope~$M\subseteq\RR^{\cardof{\setclocks}}$.
Then, two clock assignments~$\amap,\amapp\in\RR^{\cardof{\setclocks}}$ are similar if they agree on the membership in $M$.
That is,~$\amap$ and~$\amapp$ are similar if~$\amap, \amapp\in M$ or $\amap,\amapp\notin M$.
The polytope $M$ we consider will stem from the transition guards in~$\anaut$.
Similarity thus means that~$\anaut$ cannot distinguish the two clock assignments---they fail for the same reason.

Clock similarity naturally extends to sets of polytopes.
The set of polytopes along which we differentiate clock assignments is taken from a \emph{proof template}.
A proof template for a trace is a unique Hoare proof where placeholders are used instead of actual time progressions.
Hence, the explanations under consideration are instances of the template, i.e., can be obtained by replacing the placeholders with the appropriate time progressions.
More importantly, the template gives rise to a set of \emph{atomic constraints} from which all polytopes appearing in the explanations can be constructed (using Boolean connectives).
Overall, this means that two explanations are similar if the clocks they allow for are similar wrt. the polytopes of the associated proof template, meaning that~$\anaut$ cannot distinguish them and thus fails for the same reason.
\techrep{We make this precise.}

A proof template for events~$\anevent_1\dots\anevent_k$ is a Hoare proof of the form \[
	\set{\sinit} \, \atimevar_0 \, \cdots \, \set{\aprop_{2i-1}} \, \anevent_i \, \set{\aprop_{2i}} \, \atimevar_i \, \set{\aprop_{2i+1}} \, \cdots \, \atimevar_k \, \set{\false}
	\enspace.
\]
This proof is a template because $\vecof{\atimevar}=\atimevar_0,\dots,\atimevar_k$ are symbolic time progressions, i.e, they can be thought of as variables rather than actual values from $\RRplus$.
An instance of the template is a valid Hoare proof \[
	\set{\sinit} \, \atime_0 \, \cdots \, \set{\apropp_{2i-1}} \, \anevent_i \, \set{\apropp_{2i}} \, \atime_i \, \set{\apropp_{2i+1}} \, \cdots \, \atime_k \, \set{\false}
\]
with actual time progressions~$\vecof{\atime}=\atime_0,\dots,\atime_k$ such that the $\aprop_i$ subsume the $\apropp_i$ for the given choice of symbolic time progressions, $\apropp_i\prel\aprop_i[\vecof{\atimevar}\mapsto\vecof{\atime}]$.

For the classification to work, we require the following properties of~templates:
\begin{compactitem}
	\item[~~{\tagx[C1]{prop:classification:unique}}] the template is uniquely defined by the sequence $\atimevar_0.\anevent_1\dots\anevent_k.\atimevar_k$, and
	\item[~~{\tagx[C2]{prop:classification:qfree}}] the symbolic configurations appearing in the $\aprop_i$ are quantifier-free.
\end{compactitem}
The former property associates a unique template to every trace.
This is necessary for a meaningful classification via templates.
The latter property ensures that the atomic constraints we extract from the template (see below) will contain only clocks from $\setclocks$.
This is necessary for equisatisfiability to be meaningful.
In \Cref{sec:hoare-theory} we show that weakest preconditions generate appropriate templates.

An atomic clock constraint is an (in)equality over symbolic time progressions and ordinary clocks (from $\setclocks$).
We write $\tconsof{\aprop}$ for all such constraints syntactically occurring in $\aprop$.
For $\aprop_i$ from the above proof template, $\tconsof{\aprop_i}$ is a set of building blocks from which the $\apropp_i$ of \emph{all} instantiations can be constructed.
Moreover, $\anaut$ cannot distinguish time progression beyond $\tconsof{\aprop_i}$, making them ideal candidates for judging similarity.

We turn to the definition of the equivalence relation~$\erel$.
To that end, consider two explanations~$\anexp,\anexpp$ of the following form
\begin{align*}
	\anexp\text{:}\qquad
	\set{\sinit} \, \cdots \, \set{\apropp_{2i-1}} \, \anevent_i \, \set{\apropp_{2i}} \, &\atime_i \,\set{\apropp_{2i+1}} \, \cdots \, \set{\false}
	\\
	\anexpp\text{:}\qquad
	\set{\sinit} \, \cdots \, \set{\apropp'_{2i-1}} \, \anevent_i \, \set{\apropp'_{2i}} \, &\atime'_i \,\set{\apropp'_{2i+1}} \, \cdots \, \set{\false}
	\:.
\intertext{
	The events~$\anevent_1,\dots,\anevent_k$ match in both explanations, but the time progressions $\vecof{\atime}$ and $\vecof{\atime'}$ may differ.
	(Explanations with distinct event sequences are never related by~$\erel$.)
	Both explanations are instances of the same proof template~$\anout$,
}
	\;\anout\text{:}\qquad
	\set{\sinit} \, \cdots \, \set{\aprop_{2i-1}} \, \anevent_i \, \set{\aprop_{2i}} \, &\atimevar_i \,\set{\aprop_{2i+1}} \, \cdots \, \set{\false}
	\:.
	&&
\end{align*}
Now, for~$\anexp$ and~$\anexpp$ to be similar,~$\anexp\erel\anexpp$, we require the~$\apropp_i$ and~$\apropp'_i$ to satisfy the exact same atomic clock constraints appearing in~$\aprop_i$ relative to the appropriate instantiation of the symbolic clock values.
It is worth stressing that we require satisfiability, not logical equivalence, because we want the clocks to be similar, not equal.
We write~$\issatof{\aform}$ if~$\aform$ is satisfiable, that is, if there is an assignment~$\amap$ to the free variables in~$\aform$ such that~$\amap\models\aform$.
Formally then, we have:
\[
	\anexp\erel\anexpp
	\qquad\text{iff}\qquad
	\forall i~
	\forall \aform\prall{\in}\tconsof{\aprop_i}.~~
	\issatof{\aform[\vecof{\atimevar}\mapsto\vecof{\atime}]}
	\prall{\iff}
	\issatof{\aform[\vecof{\atimevar}\mapsto\vecof{\atime'}]}
	\ .
\]
It is readily checked that $\erel$ is an equivalence relation, that is, is reflexive, symmetric, and transitive, as alluded to in the beginning.
Transitivity, in particular, is desirable in our use case.
First, it means that all explanations from a class~$\awitset_i$ of~$\awitset$ are pairwise similar, that is, exhibit the same fault.
Second, the partitions are guaranteed to be disjoint.
Finally, it allows for the partitioning of~$\awitset$ to be computed efficiently (by tabulating the result of the $\issat$ queries), provided the $\issat$ queries are efficient for the type of (in)equalities used.
\begin{lemma}
	\label{thm:classification-equivalence}
	Relation $\erel$ is an equivalence relation.
\end{lemma}

\medskip\noindent
\begin{minipage}{\textwidth}
	\begin{minipage}{.563\textwidth}
		\begin{example}
			\label{ex:classification}
			\tighten
			We classify the explanations of \code{Test-1} and \code{Test-2}, which correspond to the proofs from \Cref{fig:example-proofs} with the \code{log} events removed (cf. \Cref{ex:explanation}).
			Both explanations agree on the sequence of events.
			\Cref{fig:example-classification} gives their common template.
			The atomic clock constraints are $\atimevar_1 + \atimevar_2 < 50$, $\exClock + \atimevar_1 < 50$, and $\exClock + \atimevar_1 + \atimevar_2 < 50$.
			\code{Test-1} and \code{Test-2} are similar because each clock constraint is satisfiable after instantiating the symbolic time progressions with the values in the respective trace.
			Hence, our classification groups these explanations together, \code{Test-1}$\,\erel\,$\code{Test-2}.%
		\end{example}
	\end{minipage}
	\hfill
	\begin{minipage}{4.95cm}
		\input{content/examples/classification}
	\end{minipage}
\end{minipage}

%% file: content/examples/classification.tex

\begin{figure}[H]
	\vspace{-8mm}
	\addtolength{\jot}{-0.2em}
	\newcommand{\sconf}[3]{(\astate_{#1},\astate_{#2}:{#3})}
	\newcommand{\sconfal}[3]{(\astate_{#1},\astate_{#2}&:{#3})}
	\newcommand{\clk}{\exClock}
	\newcommand{\ctx}{\exMode}
	\renewcommand{\set}[1]{{\textcolor{ProofColor}{\{#1\}}}}
	\centering
    \begin{tracebox}{Template<Test-1, Test-2>\hfill\smash{\classA}}
\begin{lstlisting}[mathescape, escapeinside={(*}{*)}]
$\set{\sconf{1}{4}{\atimevar_1 \prall{+} \atimevar_2 \prall{<} 50}}$
res CTR ack 5
$\set{\sconf{0}{4}{\ctx \prall{\neq} 0 \land \clk \prall{+} \atimevar_1 \prall{+} \atimevar_2 \prall{<} 50}}$
[ $\atimevar_2$ms] 
$\set{\sconf{0}{4}{\ctx \prall{\neq} 0 \land \clk \prall{+} \atimevar_1 \prall{<} 50}}$
req CTR get
$\set{\sconf{2}{4}{\ctx \prall{\neq} 0 \land \clk \prall{+} \atimevar_1 \prall{<} 50}}$
[ $\atimevar_1$ms] 
$\set{\sconf{2}{4}{\ctx \prall{\neq} 0}}$
res CTR ret 0
$\set{\false}$
\end{lstlisting}
	\end{tracebox}
	\vspace{-5pt}
	\caption{%
		Proof template for the explanations of \code{Test-1} and \code{Test-2}.%
		\label{fig:example-classification}%
	}
\end{figure}

%% file: content/hoare.tex

\section{Hoare Proofs with Timing}
\label{sec:hoare-theory}

For the techniques presented so far to be useful, it remains to construct Hoare proofs for traces $\aword$.
Strongest postconditions and weakest preconditions are the standard way of doing so.
The former yields efficient fault localization (\Cref{sec:localization}).
The latter satisfies the requirements for templates (\Cref{sec:classification}).
Moreover, interpolation between the two produces concise proofs beneficial for fault explanations (\Cref{sec:explanation}).

It is worth pointing out that the aforementioned concepts are well-understood for programs and ordinary automata.
However, they have not been generalized to a setting like ours where timing plays a role.
Indeed, works like \cite{1702875,DBLP:conf/rex/SchneiderBM91,DBLP:journals/fac/Hooman94,DBLP:conf/tacas/HaslbeckN18} involve timing, but do not develop the Hoare theory required here.
\techrep{The remainder of this section is devoted to develop that theory.}

\input{content/hoare_post}
\input{content/hoare_pre}
\input{content/hoare_proofs}

%% file: content/hoare_post.tex

\smartsec{Strongest Postconditions}
We compute the \emph{post image}, that is, make precise how $\anaut$ takes steps from symbolic configurations.
A step from a symbolic configuration $(\astate,\aform)$ due to transition $\astate\transofNEW{\timeevent}{\aguard}{\anup}\astatep$ on time progression $\atime$ can be taken if the guard is enabled after waiting for $\atime$ time.
After waiting, all clocks $\aclock$ are $\aclockp = \aclock + \atime$.
This means before waiting we have $\aclock = \aclockp - \atime$.
However, clocks are always non-negative, $\aclockp - \atime \geq 0$.
Overall, we replace in $\aform$ all clocks by their old versions and enforce non-negativity, $\aform'=\aform[\setclocks \mapsto \setclocks - \atime] \land \setclocks\geq \atime$.
It remains to check guard $\aguard$ and apply update $\anup$.
It is easy to see that the set of valuations in $\aform'$ satisfying~$\aguard$ is precisely $\aformp=\aform'\land\aguard$.
To perform a singleton update $\set{x \mapsto y}$, we capture the new valuation of $x$ by the equality $x=y$.
To avoid an influence of the update of $x$ on other variables/clocks, we have to rewrite $\aformp$ to not contain $x$.
This is needed as $\aformp$ might use $x$ to correlate other variables/clocks---we want to preserve these correlations without affecting them.
We use an existential abstraction that results in $\aformp'=\exists z.\; \aformp[x\mapsto z]\land x = y$.
Then, the post image is $(\astatep,\aformp')$.
For stuttering steps, we add the original configuration $(\astate,\aform)$ to the post image.
Steps due to events from $\setevents$ are similar.

We define a symbolic transformer that implements the above update of the symbolic encoding $\aform$ to $\aformp'$ in the general case:
\[
	\ssemof{\aguard}{\setcompact{\vecof{x}\mapsto\vecof{y}}}{\atime}(\aform)
	~\defeq~
	\exists \vecof{z}.~
	(\aform[\setclocks\mapsto\setclocks\prall{-}\atime\mkern+1mu] \land \setclocks\geq\atime \land \aguard)
	[\mkern+1mu\vecof{x}\mapsto\vecof{z}\mkern+1mu]
	\wedge
	\vecof{x}=\vecof{y}
	\enspace,
\]
where $\vecof{x}$ is short for a sequence $x_1,\dots,x_m$ of variables/clocks.
We arrive at:
\begin{align*}
	\postof{\atime}{\aprop}
		&\defeq
		\setcond{
			(\astatep,\ssemof{\aguard}{\anup}{\atime}(\aform))
		\!\!}{\!\!
			(\astate,\aform)\in\aprop \wedge
			\astate\transofNEW{\timeevent}{\aguard}{\anup}\astatep
		}
		\,\cup\; \bigl(\atime = 0 ~?~ \aprop \::\: \emptyset\bigr)
	\\
	\postof{\anevent}{\aprop}
		&\defeq
		\setcond{
			(\astatep,\ssemof{\aguard}{\anup}{0}(\aform))
		\!\!}{\!\!
			(\astate,\aform)\in\aprop \wedge
			\astate\transofNEW{\anevent}{\aguard}{\anup}\astatep
		}
		\ .
\end{align*}
The post image is sound and precise in the sense that it captures accurately the steps the configurations denoted by $\aprop$ can take.
The \namecref{thm:soundness-abstract-post} makes this precise.

\begin{lemma}
	\label{thm:soundness-abstract-post}
	$\aconfp \models\postof{\asymbol}{\aprop}$ iff there is $\aconf \models \aprop$ with $\aconf\stepofNEW{\asymbol}\aconfp$.
\end{lemma}

\begin{example}
	\label{ex:post}
	We apply $\post{}$ to $\aprop = \setcompact{(\astate_4, 49 \leq \exClock \leq 52)}$ for $\exAutDog$ from \Cref{fig:example-aut}.
	Recall that $\exAutDog$ resets variable $\exMode$ within $5\milis$ after $\exClock$ has reached the $50\milis$ mark.
	Indeed, $\postof{1}{\aprop}$ for $1\milis$ contains both the resetting and the non-resetting case: $(\astate_5, {50 \leq \exClock \leq 53} \land {\exMode = 0})$ and $(\astate_4, 50 \leq \exClock \leq 53)$.

	The post image still lacks a way to commute with the \labelcref{eq:trace-equalities} congruences.
	While $\postof{5}{\postof{1}{\aprop}}$ witnesses the reset via condition ${55 \leq \exClock \leq 58} \land {\exMode = 0}$ for both $\astate_4$ and $\astate_5$, it is not equivalent to $\postof{6}{\aprop}$, which is $\false$ since all transitions in $\astate_4$ are disabled for a full $6\milis$ wait.
\end{example}

While the post image captures the individual steps of basic runs on traces~$\aword$, we have to consider the basic runs of all traces $\awordp \tequiv \aword$ to generate a Hoare proof for $\aword$.
Basically, the \eqref{eq:trace-equalities} equivalences state that the time progressions between events can be split/merged arbitrarily.
To that end, we define the strongest postcondition $\spost$ which inspects all basic runs simultaneously, intuitively, by rewriting according to the \eqref{eq:trace-equalities} equivalences on-the-fly.
(Note that normalization according to \Cref{sec:model} avoids the merging case of \eqref{eq:trace-equalities}.)
Then, for events $\anevent$ the strongest postcondition merely applies the post image to $\anevent$.
For time progressions $\atime$, the strongest postcondition considers all decompositions of $\atime$ into fragments $\atime_1,\dots,\atime_k$ that add up to $\atime$ and applies the post image iteratively to all the $\atime_i$.
This includes stuttering where 0 is rewritten to 0\dots0.
If there are loops in $\anaut$, the strongest postcondition might need to consider infinitely many decompositions.
We address this problem by enumerating decompositions of increasing length and applying to each decomposition a widening $\widening$ with the following properties:
\begin{inparaenum}
	\item the result of the widening is $\prel$-weaker than its input, $\aprop_i\prel\widening(\aprop_1,{\cdots},\aprop_k)$ for all $i$, and
	\item the widening stabilizes after finitely many iterations for $\prel$-increasing sequences, $\aprop_1\prel\aprop_2\prel{\cdots}$ implies that there is $k$ so that $\widening(\aprop_1,{\cdots},\aprop_i)=\widening(\aprop_1,{\cdots},\aprop_{i+1})$ for all $i\geq k$.
\end{inparaenum}
We write $\widening(\aprop_i)_{i\in\nat}$ and mean the stabilized $\widening(\aprop_1,{\cdots},\aprop_k)$.
Given a widening, the strongest postcondition is:
\begin{mathpar}
	\spost[\atime](\aprop)
		\defeq
		\widening\left(
				\exists \atime_1,{\cdots},\atime_i.~
				\atime=\atime_1+\cdots+\atime_i
				~\wedge~
				\post{\atime_i}\circ\cdots\circ\post{\atime_1}(\aprop)
		\right)_{i\in\nat}
	\and
	\spost[\anevent](\aprop)
		\defeq
		\post{\anevent}(\aprop)
	\and
	\spost[\asymbol.\aword](\aprop)
		\defeq
		\spost[\aword] \circ \spost[\asymbol](\aprop)
	\and
	\spof{\aprop}{\aword}
		\defeq
		\spost[\aword](\aprop)
\end{mathpar}
where the $\atime_1,\dots,\atime_i$ are fresh.
Observe that the sequence of post images in $\spost[\atime]$ is $\prel$-increasing:
one can always extend the decomposition by additionally waiting for $0$ time, $\post{\atime_i}(\aprop)\prel\post{0}\circ\post{\atime_i}(\aprop)$.
The strongest postcondition considers all basic runs and $\widening$ overapproximates the reachable configurations.
It is sound.

\begin{lemma}
	\label{thm:sp-valid-hoare}
	If $\spof{\aprop}{\aword}\prel\apropp$, then $\models\hoare{\aprop}{\aword}{\apropp}$.
\end{lemma}

For the \namecref{thm:sp-valid-hoare} to be useful, an appropriate widening $\widening$ is required.
In general, finding such a widening is challenging---after all, it resembles finding loop invariants---and for doing so we refer to existing works, like~\cite{DBLP:conf/oopsla/DilligDLM13,DBLP:conf/fm/FlanaganL01,DBLP:conf/popl/CousotH78}, to name a few.
In practice, a widening may be obtained more easily.
In case $\anaut$ is free from $\timeevent$-cycles, stabilization is guaranteed after $k$ iterations, where $k$ is the length of the longest simple $\timeevent$-path.
If there are $\timeevent$-cycles, stabilization is still guaranteed after $k$ iterations if all $\timeevent$-cycles are \emph{idempotent}.
A $\timeevent$-cycle is idempotent if repeated executions of the cycle produce only configurations that already a single execution of the cycle produces.
Interestingly, idempotency can be checked while computing the widening: if the $(k\prall{+}1)$st iteration produces new configurations, idempotency does not hold.
In our setting, idempotency was always satisfied.
For the remainder of this paper, we assume an appropriate widening is given.

%% file: content/hoare_pre.tex

\smartsec{Weakest Preconditions}
We also compute weakest preconditions, the time-reversed dual of strongest postconditions.
Our definition will satisfy the template requirements \eqref{prop:classification:unique} and \eqref{prop:classification:qfree} from \Cref{sec:classification}.

\tighten
The \emph{pre image} is the set of symbolic configurations that reach a given configuration in automaton $\anaut$.
Consider some $(\astatep,\aformp)$ and $\astate\transofNEW{\timeevent}{\aguard}{\anup}\astatep$.
The pre image first rewinds updates $\anup\prall{=}\set{\vecof{x}\mapsto \vecof{y}}$ by replacing $\vecof{x}$ with $\vecof{y}$.
Then, it adds a disjunct $\aformpp \prall{=} \aformp[\vecof{x} \mapsto \vecof{y}] \lor \neg\aguard$.
Adding the disjunct makes the pre image weaker; it does not affect soundness in \Cref{thm:soundness-abstract-pre} which ignores the \emph{stuck} configurations denoted by $(\astate,\neg\aguard)$.
Finally, we rewind the clock progression~$\atime$ by replacing all clocks $\aclock$ in $\aformpp$ with $\aclock+\atime$.
We arrive at the pre image $\aform = \aformpp[\setclocks\mapsto\setclocks + \atime]$.
Transitions due to events are similar.
We define a symbolic transformer to apply the above process: \[
	\smesof{\aguard}{\setcompact{\vecof{x}\mapsto\vecof{y}}}{\atime}(\aformp)
	\,\defeq\,
	(\aformp[\vecof{x}\mapsto\vecof{y}] \lor \lnot \aguard)[\setclocks \mapsto \setclocks + \atime]
	\ .
\]

To account for other transitions leaving $\astate$ that are enabled in $\aformpp$, we compute the meet $\meet$ of the per-transition pre images.
Intuitively, this intersects symbolic configurations on a per-state basis, ensuring that any configuration from the pre image either gets stuck or steps to one of the configurations we computed the pre image for.
Technically, the meet $\meet$ for sets $M$ of symbolic configurations~is:
\[
	\bigmeet M \defeq \setcond{ (\astate, {\textstyle\bigwedge_{(\astate, \aform) \in M}}\; \aform)}{\astate \in \setstates}
	\ .
\]
Notably, when considering the meet of $M$, we cannot understand $M$ as a condition.
This is because conditions treat symbolic configurations disjunctively and can be normalized by \Cref{thm:prop-rewriting}.
However, the meet is not preserved under these transformations.
We write $M_1\meet M_2$ to mean $\bigmeet(M_1\cup M_2)$.

The discussion yields the following definition of the pre image:
\begin{align*}
	\preof{\atime}{\aprop}
	&\defeq
		\bigmeet\setcond{\!
		(\astate,\smesof{\aguard}{\anup}{\atime}(\aformp))
	\!\!}{\!\!
		(\astatep,\aformp)\in\aprop \land
		\astate\transofNEW{\mkern-1mu\timeevent}{\mkern-1mu\aguard}{\mkern-1mu\anup}\astatep
	\!}
	\,\,{\meet}\; \bigl(\atime = 0 ~?~ \aprop \::\: \emptyset\bigr)
	\\
	\preof{\anevent}{\aprop}
		&\defeq
		\bigmeet\setcond{\!
			(\astate,\smesof{\aguard}{\anup}{0}(\aformp))
		\!\!}{\!\!
			(\astatep,\aformp)\in\aprop
			\land
			\astate\transofNEW{\anevent}{\mkern-1mu\aguard}{\mkern-1mu\anup}\astatep
		\!}
	\enspace,
\end{align*}
capturing precisely the forced reachability in $\anaut$, as stated by the next \namecref{thm:soundness-abstract-pre}.

\begin{lemma}
	\label{thm:soundness-abstract-pre}
	${\aconf\models\preof{\asymbol}{\aprop}}$ iff for all $\aconfp$, $\aconf\stepofNEW{\asymbol}\aconfp$ implies $\aconfp \models \aprop$.
\end{lemma}

\begin{example}
	\label{ex:pre}
	We apply $\pre{}$ to $\aprop = \setcompact{(\astate_4, 49 \leq \exClock \leq 52)}$ for $\exAutDog$ from \Cref{fig:example-aut}.
	Computing $\preof{1}{\aprop}$ highlights the need for the meet.
	The $\Delta$-loop on $\astate_4$ does not give $(\astate_4, 48 \leq \exClock \leq 51)$ as precondition.
	Instead, it is $(\astate_4, 48 \leq \exClock < 49)$ which is the result of ${\setcompact{(\astate_4, 48 \leq \exClock \leq 51)}} \meet \setcompact{(\astate_4, \exClock \geq 54 \lor \exClock < 49)}$.
	Indeed, $\exAutDog$ reaches a non-$\aprop$ configuration via the resetting transition to $\astate_5$ if $\exClock=49$.
\end{example}

The weakest precondition $\wpof{\asymbol}{\apropp}$ denotes all configurations that either step to $\apropp$ under $\asymbol$ or have no step at all.
Technically, the weakest precondition repeatedly applies the pre image for all decompositions of time progressions.
For termination, we again rely on the widening $\widening$.
Since the pre image sequence is $\prel$-decreasing, we turn it into an increasing sequence by taking complements.
More precisely, we use the widening $\narrowing(\aprop_1,\cdots,\aprop_m)\defeq\neg\widening(\neg\aprop_1,{\cdots},\neg\aprop_m)$.
The weakest precondition is defined by:
\begin{mathpar}
	\wpre[\atime](\aprop)
	\,\defeq~
	\narrowing\left(
			\forall \atime_1,{\cdots},\atime_i.~
			\atime=\atime_1+\cdots+\atime_i
			\implies
			\pre{\atime_1}\circ\cdots\circ\pre{\atime_i}(\aprop)
	\right)_{i\in\nat}
	\and
	\wpre[\anevent](\aprop)
	\,\defeq~
	\pre{\anevent}(\aprop)
	\and
	\wpre[\aword.\asymbol](\aprop)
	\,\defeq~
	\wpre[\aword]\circ\wpre[\asymbol](\aprop)
	\and
	\wpof{\aword}{\aprop} \,\defeq~ \wpre[\aword](\aprop)
	\enspace.
\end{mathpar}
Note that $\wpre[\atime]$ applies to ordinary time progressions $\atime$ as well as symbolic time progressions $\atimevar$ appearing in proof templates.
The weakest precondition is sound\moreless{ in that it produces valid Hoare triples}{}.

\begin{lemma}
	\label{thm:wp-valid-hoare}
	If $\aprop\prel\wpof{\aword}{\apropp}$, then $\models\hoare{\aprop}{\aword}{\apropp}$.
\end{lemma}

%% file: content/hoare_proofs.tex

\smartsec{Concise Hoare Proofs}
The developed theory allows for an efficient way to produce concise Hoare proofs.
We first apply strongest postconditions to generate an initial proof.
Then, starting from the back, we apply weakest preconditions and interpolation~\cite{DBLP:journals/jsyml/Craig57} to simplify the initial proof.
We make this precise.

Combining \Cref{thm:hoare-characterizes-runs,thm:sp-valid-hoare} gives an effective way of finding faults in traces $\aword=\asymbol_1\dots\asymbol_n$ and extracting a witness:
iteratively compute the strongest postcondition for increasing prefixes of $\aword$ and check if the result is unsatisfiable.
That is, compute $\aprop=\spof{\asymbol_1.\cdots.\asymbol_k}{\sinit}$ and check if $\aprop=\false$.
If so, then $\awitness=\asymbol_1\dots\asymbol_k$ is a witness for fault $\asymbol_k$.
Otherwise, continue with the prefix $\asymbol_1\dots\asymbol_k.\asymbol_{k+1}$ which can reuse the previously computed $\aprop$: $\spof{\asymbol_1\dots\asymbol_k.\asymbol_{k+1}}{\sinit}=\spof{\asymbol_{k+1}}{\aprop}$.
As per \Cref{thm:sp-valid-hoare}, the approach gives rise to the valid Hoare proof
\[
	\set{\sinit}\; \asymbol_1 \;\cdots \;\set{\aprop_{i}}\; \asymbol_{i+1} \;\set{\aprop_{i+1}}\; \cdots\; \asymbol_k \;\set{\false}
	\quad\text{with}\quad
	\aprop_{i+1}=\spof{\aprop_i}{\asymbol_{i+1}}
	\ .
\]

It is well-known that strongest postconditions produce unnecessarily complex proofs~\cite{DBLP:reference/mc/McMillan18}.
To alleviate this weakness, we use interpolation~\cite{DBLP:journals/jsyml/Craig57}.
For two formulas $\aform$ and $\aformp$ with $\aform{\implies}\aformp$, an interpolant is a formula $\aninterpolant$ with $\aform{\implies}\aninterpolant$ and $\aninterpolant{\implies}\aformp$.
The interpolant for conditions $\aprop$ and $\apropp$ with $\aprop\prel\apropp$, denoted $\interpolantof{\aprop}{\apropp}$, results from interpolating the symbolic configurations in $\aprop$ with the corresponding ones in $\apropp$.
Interpolants exist in first-order predicate logic~\cite{DBLP:journals/jsyml/Craig57,Lyndon1959AnIT}.

From the above $\spost{}$- generated proof we construct an interpolated proof \[
	\set{\sinit}\; \asymbol_1 \;\cdots \;\set{\interpolantof{\aprop_{i}}{\apropp_{i}}}\; \asymbol_{i+1} \;\set{\interpolantof{\aprop_{i+1}}{\apropp_{i+1}}}\; \cdots\; \asymbol_k \;\set{\false}
\] using $\wpre{}$ as follows.
Assume we already constructed, starting from the back, the interpolants $\interpolantof{\aprop_k}{\apropp_k}$ through $\interpolantof{\aprop_{i+1}}{\apropp_{i+1}}$.
Now, the goal is to obtain an interpolant $\aninterpolant$ so that $\hoare{\aninterpolant}{\asymbol_{i+1}}{\interpolantof{\aprop_{i+1}}{\apropp_{i+1}}}$ is valid.
The weakest precondition for the latest interpolant yields $\apropp_i=\wpof{\asymbol_i}{\interpolantof{\aprop_{i+1}}{\apropp_{i+1}}}$.
This gives a valid Hoare triple $\models\hoare{\apropp_i}{\asymbol_{i+1}}{\interpolantof{\aprop_{i+1}}{\apropp_{i+1}}}$.
Our goal is to interpolate $\aprop_i$ and $\apropp_i$.
If $\aprop_i\prel\apropp_i$, we can interpolate $\aprop_i$ and $\apropp_i$ to obtain $\aninterpolant=\interpolantof{\aprop_i}{\apropp_i}$.\footnote{
	One can show that the inclusion $\aprop_i\prel\apropp_i$ is always satisfied in our setting where $\timeevent$-cycles are idempotent and the widenings $\widening$ and $\narrowing$ simply enumerate all necessary decompositions of time progressions.
	\moreless{\Cref{app:interpolation-guarantee} gives}{Refer to \cite{techreport} for} a more general property.
}
Otherwise, we simply choose $\aninterpolant=\apropp_i$.
By \Cref{thm:wp-valid-hoare} together with $\aninterpolant\prel\apropp_i$, we know that $\models\hoare{\aninterpolant}{\asymbol_{i+1}}{\interpolantof{\aprop_{i+1}}{\apropp_{i+1}}}$ is valid.
Overall, this constructs a valid proof.

%% file: content/eval.tex

\section{Application in Automotive Software}
\label{sec:iav}

We implemented and tested our approach on benchmarks provided by our project partner from the automotive industry.
The implementation parses, classifies, and annotates traces of ECUs running the Unified Diagnostic Services (UDS).
We turned a PRD with 350 pages of natural language specifying $23$ services into a PRD automaton of  12.5k states and 70k transitions.
We evaluated our tool on $1000$ traces which are processed within $24$ minutes.
Our tool is implemented in \code{C\#} and processes traces in the three stages explained below.
It naturally supports multi-threading for the localization, explanation, and classification since they are agnostic to the (set of) other traces being analyzed.

\smartsec{Preprocessing Stage}
The first stage parses trace files and brings them into a shape similar to \Cref{fig:example-tests}.
UDS specify a request-response protocol for ECUs communicating over a CAN bus.
The traces are a recording of all messages seen on the bus during a test run.
We found the preprocessing more difficult than expected, because the trace files have a non-standard format.
These problems stem from the fact that our industrial partner creates tests partly manually and inserts natural language annotations.
A useful type of annotation that we could extract are the positions deemed erroneous by the test environment.

\smartsec{Modeling Stage}
The second stage creates the test model, a PRD automaton as defined in \Cref{sec:model}.
Modeling a natural language PRD is a non-trivial and time-consuming process.
To translate the PRD into an automaton, we developed an API capable of programmatically describing services and their communication requirements.
The API supports a declarative formulation of the communication requirements which it compiles down into an automaton. 
The compilation is controlled by a set of parameters because the PRD prescribes different behavior depending on the ECU version (and related static parameters).
There are further high-level modeling constructs such as regular expressions, as alluded to in \Cref{sec:model} and seen in \Cref{fig:example-aut}.

Unfortunately, not all requirements from the PRD are restricted to the trace: they may refer to events internal to the ECU that are not contained in the trace files.
While our API and PRD automata are capable of expressing these requirements, the test environment is unable to detect them.
To circumvent the problem of missing information, we over-approximated our model using non-determinism.
That is, we simply allow our model to do any of the specified behaviors for unobservable internal events.
A downside of this is that errors dependent on these events cannot be found during fault localization.

\smartsec{Analysis Stage}
The last stage performs fault localization~(\cref{sec:localization}), explanation~(\cref{sec:explanation}), and classification~(\cref{sec:classification}).
We carefully inspected 86 traces curated by our industrial partner.
The tests targeted one of the 23 services, yet they contain requests and responses to a multitude of services responsible for setting up the ECU configuration.
The annotations of the test environment marked 100 faults, 95 of which are also found by our fault localization.
Our tool finds and explains another 10 undetected faults, which totals to 105 fault explanations.
The five faults missed by our localization are actually incorrect annotations by the test environment, which we will explain in a moment.

\Cref{fig:statistics} gives the lengths of the found witnesses and the average lengths of their explanations.
The explanation lengths are closely tied to the kinds of faults in the test set.
In our set, long witnesses tend to have a long prefix unimportant to the fault.
This is reflected in the partitioning found by our classification.
%
\input{content/stats.tex}

The classification divides the faults into six partitions.
We found that each partition belongs to one of the following three error types:
\begin{inparaenum}
    \item ECU responds too late (1+8); 
    \item ECU fails to reset a variable upon restart (2); 
    \item ECU responds when it should not (2+1+91).
\end{inparaenum} 
Here, 1+8 means we have two partitions and one with a single witness, one with eight equivalent witnesses.
Each error type consists of at most two relevant events.
Unrelated events in-between those two events are dropped by fault explanation.
The relevant events are:
\begin{inparaenum}
    \item the request and the late response,
    \item the response event which revealed that the variable has not been reset, and
    \item the request and the incorrectly given response.
\end{inparaenum}

There are two partitions with error type~(i).
This is because the late response is given by another service and thus leads to different control flow in the automaton.
Indeed, there might be distinct root causes: different services are likely controlled by different pieces of code.
A similar reason produces three partitions of error type (iii).
Interestingly, the singleton partition for~(i) is completely missed by the test environment (no fault was marked).
This supports our claim that the test environment only detects faults targeted by the tests and ignores other faults.
The other partition of (i) was detected by the test environment by accident:
in some traces, the ECU response is so late that the test environment incorrectly marks the response as missing.
These incorrect marks represent no faults and are not considered by our localization.
Instead, our localization actually detects the late responses and marks them correctly. 

Our tool provides a partitioning file with direct links to the trace files.
It also modifies the trace files to highlight the events related to the fault (cf. \cref{sec:explanation}) and provides an intuitive explanation of the fault. 
As for the latter, the user is informed about the difference between the observed and the automaton-expected behavior.
Our manual inspection showed no incorrect classification.
That is, our tool has never grouped together traces which test engineers would deem caused by distinct faults.
This is promising feedback because incorrect classification is dreaded: a single missed flaw of an ECU can cause large costs.
Overall, we reduced the workload of manually inspecting 86 traces with 100 fault marks to inspecting six representative faults that expose more misbehavior than marked by the test environment.

%% file: content/stats.tex

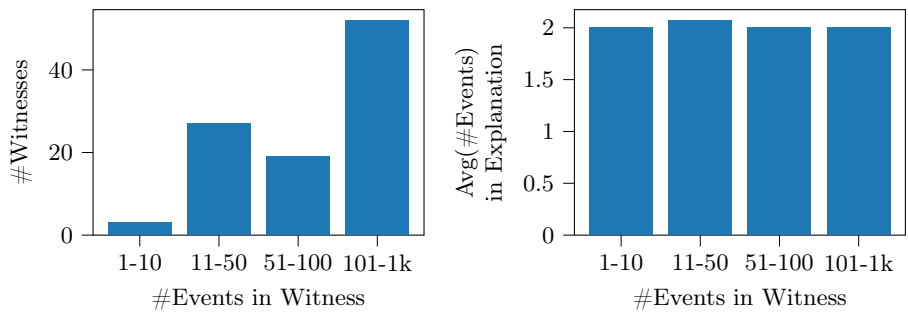
\begin{figure}[t]
    \begin{tikzpicture}
        \definecolor{darkgray}{RGB}{176,176,176}
        \definecolor{steelblue}{RGB}{31,119,180}
        \begin{groupplot}[
            scale only axis,
            width=1/2*\textwidth-1.7cm,
            height=3cm,
            group style={group size=2 by 1, horizontal sep=2cm},
            tick align=outside,
            tick pos=left,
            x grid style={darkgray},
            xtick style={color=black},
            xtick={0,1,2,3},
            xticklabels={1-10,11-50,51-100,101-1k},
            y grid style={darkgray},
            ytick style={color=black},
            xlabel={\#Events in Witness},
            xmin=-0.59, xmax=3.59,
        ]

            \nextgroupplot[
                ylabel={\#Witnesses},
                ymin=0, ymax=54.6,
            ]
            \draw[draw=none,fill=steelblue] (axis cs:-0.4,0) rectangle (axis cs:0.4,3);
            \draw[draw=none,fill=steelblue] (axis cs:0.6,0) rectangle (axis cs:1.4,27);
            \draw[draw=none,fill=steelblue] (axis cs:1.6,0) rectangle (axis cs:2.4,19);
            \draw[draw=none,fill=steelblue] (axis cs:2.6,0) rectangle (axis cs:3.4,52);

            \nextgroupplot[
                align=center,
                ylabel={Avg(\#Events)\\ in Explanation},
                ymin=0, ymax=2.175,
            ]
            \draw[draw=none,fill=steelblue] (axis cs:-0.4,0) rectangle (axis cs:0.4,2);
            \draw[draw=none,fill=steelblue] (axis cs:0.6,0) rectangle (axis cs:1.4,2.07142857142857);
            \draw[draw=none,fill=steelblue] (axis cs:1.6,0) rectangle (axis cs:2.4,2);
            \draw[draw=none,fill=steelblue] (axis cs:2.6,0) rectangle (axis cs:3.4,2);
        \end{groupplot}
    \end{tikzpicture}
    \vspace{-3mm}
    \caption{%
        Statistics on witnesses: number (left) and average explanation length (right).%
        \label{fig:statistics}%
    }
    \vspace{-2.5mm}
\end{figure}

%% file: content/related.tex

\section{Related Work}
\label{sec:related}

\smartsec{Fault Explanation}
Our work on fault explanation is related to minimizing unit tests in~\cite{Zeller07}: tests are pruned by removing the commands that are not contained in a backward slice from a failing instruction.
With timing constraints, slicing does not work (every command is relevant), which is why we have developed our approach based on Hoare logic.
The assertions provided by a Hoare proof have the additional advantage of being able to prune even dependent commands inside a slice (based on the relationship between intermediary assertions), which leads to higher reduction rates.
Similar to our approach is the fault localization and explanation from~\cite{DBLP:conf/vmcai/ChristESW13,DBLP:conf/icse/Schwartz-Narbonne15}.
That work also makes use of interpolation~\cite{DBLP:conf/cav/McMillan03} and is able to strip infixes from a trace despite dependencies.
Our fault localization can be understood as a generalization to a timed setting where every command contributes to the progression of time and therefore is delicate to remove.

A popular fault explanation approach that can be found in several variants in the literature~\cite{DBLP:conf/sigsoft/Zeller02,DBLP:conf/spin/GroceV03,DBLP:conf/popl/BallNR03,DBLP:conf/kbse/RenierisR03,DBLP:conf/tacas/Groce04,Guo06,DBLP:conf/pldi/JoseM11} is Delta debugging: starting from a failing test, produce a similar but passing test, and take the difference in commands as an explanation of the fault.
In~\cite{DBLP:conf/sigsoft/Zeller02,DBLP:conf/spin/GroceV03,DBLP:conf/kbse/RenierisR03,DBLP:conf/tacas/Groce04,Guo06}, the passing test is found by repeatedly testing the concrete system~\cite{DBLP:conf/spin/GroceV03}, which is impossible in our in-vitro setting.
In~\cite{DBLP:conf/popl/BallNR03,DBLP:conf/tacas/Groce04,DBLP:conf/pldi/JoseM11}, a model checker resp. a solver is queried for a passing test resp. a satisfiable subset of clauses.
Our Hoare proof can be understood as building up an alternative and valid execution.
Different from a mere execution, however, intermediary assertions provide valuable information about the program state that we rely on when classifying tests.

The explanation from \cite{DBLP:conf/tacas/JinRS02} divides a computation into fated and free segments, the former being deterministic reactions to inputs and the latter being inputs that, if controlled appropriately, avoid the fault and hence should be considered responsible for it.
The segments are computed via rather heavy game-theoretic techniques, which would be difficult to generalize to timed systems.
A more practical variant can be found in~\cite{DBLP:conf/kbse/WangR05,DBLP:conf/icse/ZhangGG06}.
These works modify tests in a way that changes the evaluation of conditionals.
Neither can we re-run tests in an in-vitro setting, nor would we be able to influence the timing behavior.

There is a body of literatur on statistical approaches to finding program points that are particularly prone to errors, see the surveys~\cite{RegressionSurvey2012,DBLP:journals/tse/WongGLAW16}.
We need to pinpoint the precise as possible cause of a bug, instead.

\smartsec{Fault Classification}
Previous works on test case classification follow the same underlying principle~\cite{DBLP:conf/issta/GolaghaLPI19,DBLP:conf/issre/JordanHFP20,DBLP:conf/icse/GolaghaRMHP18,DBLP:conf/sigsoft/DiGiuseppeJ12a,DBLP:conf/icse/GolaghaPFN17,DBLP:conf/icse/DickinsonLP01,DBLP:conf/icse/PodgurskiLFMMSW03,DBLP:conf/sigsoft/LiuH06,DBLP:conf/fase/PhamKRR17}:
devise a distance metric on test cases that is used to group them.
The metrics are based on properties like  
the commonality/frequency of words in comments and variables in the code~\cite{DBLP:conf/sigsoft/DiGiuseppeJ12a} or the correlation of tests failing/passing in previous test runs~\cite{DBLP:conf/issta/GolaghaLPI19}.
Symbolic execution has been used to derive more semantic properties based on the source code location of faults~\cite{DBLP:conf/sigsoft/LiuH06} and the longest prefix a failing trace shares with some passing trace~\cite{DBLP:conf/fase/PhamKRR17}.
The problem is that the suggested metrics are at best vague surrogates for the underlying faults.
Using a model-based approach, we compare traces not against each other but against a ground truth (the PRD automaton).

Another related line of work is test case prioritization, test case selection, and test suite minimization~\cite{RegressionSurvey2012}.
Although formulated differently, these problems share the task of choosing tests from a predefined pool.
Experiments have shown that manually chosen test suites outperform automatically selected ones in their ability to expose bugs~\cite{Clustering09}.
To increase the number of tests that can be evaluated manually by an expert, the literature has proposed the use of clustering algorithms to group together tests with similar characteristics (so that the expert only has to evaluate clusters).
The clustering is computed from syntactic information (a bitwise comparison of test executions).
As argued before, we use semantic information and compute the classification wrt. a ground truth.

On the automatic side, \cite{DBLP:conf/tap/PodelskiSW16} suggests the use of Hoare proofs to classify error traces.
Our approach follows this idea and goes beyond it with the proposal of proof templates.
Proof templates seem to be precisely the information needed to classify tests that are subject to real-time constraints.
Harder et al. suggest to minimize test suites based on likely program invariants inferred from sample values obtained in test runs \cite{OperationalAbstraction03}.
Hoare triples are more precise than invariants, even more so as we work with a ground truth rather than sample values.

%% file: appendix/base.tex

\input{appendix/details}
\input{appendix/proofs}

%% file: appendix/details.tex

\section{Additional Material}

We provide details missing in the main part of the paper.

\subsection{Missing Definitions}
We lift quantification and the standard Boolean connectives to conditions $\aprop$, where $\aformp$ are formulas, as follows:
\begin{mathpar}
	{\exists\,\vecof{x}.\;\aprop ~\defeq~ \setcond{(\astate,\exists\,\vecof{x}.\;\aform)}{(\astate,\aform)\in\aprop}}\strut
	\and
	{\aformp\land\aprop ~\defeq~ \setcond{(\astate,\aformp\land\aform)}{(\astate,\aform)\in\aprop}}\strut
	\and
	{\aformp\Rightarrow\aprop ~\defeq~ \setcond{(\astate,\aformp\Rightarrow\aform)}{(\astate,\aform)\in\aprop}}\strut
	\and
	{\neg\aprop ~\defeq~ \setcond{(\astate,\neg\aform)}{(\astate,\aform)\in\aprop}}\strut
\end{mathpar}

For $\vecof{x}=x_1,\dots,x_m$ and $\vecof{y}=y_1,\dots,y_m$ and $\setclocks=\setcompact{\aclock_1,\dots,\aclock_m}$ we define the following abbreviations:
\begin{mathpar}
	\vecof{x}=\vecof{y} ~\defeq~ \bigwedge_{i=1}^{m} x_i=y_i
	\and
	\aformp[\vecof{x}\mapsto\vecof{z}] ~\defeq~ \aformp[x_1\mapsto z_1,\dots,x_m\mapsto z_m]
	\and
	\setclocks\geq\atime ~\defeq~ \bigwedge_{\aclock\in\setclocks} \aclock\geq\atime
	\and
	\aform[\setclocks\mapsto\setclocks\prall{-}\atime\mkern+1mu] ~\defeq~ \aform[\aclock_1\mapsto\aclock_1-\atime,\dots,\aclock_n\mapsto\aclock_n-\atime]
\end{mathpar}


\subsection{Checking Inclusion}
We devise an effective check for inclusions of the form $\aprop\prel\apropp$.
Towards such a check, consider an abstract configuration $(\astate,\aform)\in\aprop$.
Note that we cannot simply test membership of $(\astate,\aform)$ in $\apropp$ since the configurations in $\apropp$ might be strictly weaker or just syntactically different.
Instead, we test whether there is some $(\astate,\aformp)\in\apropp$ such that $\aform{\implies}\aformp$ is valid.
The implication ensures that the valuations denoted by $\aform$ are also denoted by $\aformp$.
More formally, $\amap\models\aform$ implies $\amap\models\aformp$ and thus also $(\astate,\amap)\models\aprop$ implies $(\astate,\amap)\models\apropp$, for any valuation $\amap$.
The implications can then be discharged by off-the-shelf SMT solvers like Z3~\cite{DBLP:conf/tacas/MouraB08}, which is what we use.

The inclusion check $\aprop\prel\apropp$ among conditions proposed above allows for an elegant and powerful tuning mechanism of the resulting explanation's precision.
In our experience, it might be helpful to ignore certain parts of the model $\anaut$, for instance, the timing behavior of specific clocks.
To refine the explanation in this manner, one can choose a subset $\vecof{x}\,\;{\subseteq}\:\setvars\cup\setclocks$ of variables and clocks that will be ignored during the construction of the explanation.
To ignore $\vecof{x}$, we simply adapt the above check to integrate an existential abstraction as follows: \[
	(\astate,\aform)\prel[\vecof{x}](\astate,\aformp)
	~\defiff~
	(\exists \vecof{x}.~\aform)
	{\implies}
	(\exists \vecof{x}.~\aformp)
	\ .
\]
Then, $\aprop$ is subsumed by $\apropp$ wrt. to the hidden set $\vecof{x}$, denoted by $\aprop\prel[\vecof{x}]\apropp$, if for all $(\astate,\aform)\in\aprop$ there is $(\astate,\aformp)\in\apropp$ such that $(\astate,\aform)\prel[\vecof{x}](\astate,\aformp)$.
This way, changes to the valuation of the variables/clocks in $\vecof{x}$ are ignored.
Hence, the explanation also becomes incognizant of $\vecof{x}$.

\subsection{Precision of $\spost{}$ and $\wpre{}$}
\label{app:interpolation-guarantee}
The strongest postcondition avoids false-positives in the sense that it will not produce $\false$ for any $\aword$ if $\anaut$ has a run on $\aword$.
\begin{lemma}
	\label{thm:sp-characterizes-runs}
	If $\anaut$ has a run on $\aword$, then $\spof{\sinit}{\aword}\neq\false$.
\end{lemma}

\Cref{thm:sp-valid-hoare,thm:wp-valid-hoare} state that the strongest postcondition and the weakest precondition are sound in that they generate valid Hoare triples.
The converse is also ture, provided the widenings $\widening$ and $\narrowing$ are sufficiently precise.
Formally, the widening $\widening$ (narrowing $\narrowing$) is precise, if $\widening M = \bigcup M$ ($\narrowing M = \bigmeet M$) for all~$M$.
Observe that if $\widening$ is precise, so is $\narrowing$.
\begin{lemma}
	\label{thm:precise-sp}
	If $\widening$ is precise, then $\models\hoare{\aprop}{\aword}{\apropp}$ implies $\spof{\aprop}{\aword}\prel\apropp$.
\end{lemma}
\begin{lemma}
	\label{thm:precise-wp}
	If $\narrowing$ is precise, then $\models\hoare{\aprop}{\aword}{\apropp}$ implies $\aprop\prel\wpof{\aword}{\apropp}$.
\end{lemma}

As a consequence, interpolation between strongest postconditions and weakest preconditions is possible whenever $\widening$ and $\narrowing$ are precise.
\begin{corollary}
	\label{thm:precise-sp-wp}
	If $\widening$ is precise, then $\spof{\aprop}{\aword}\prel\apropp$ iff $\aprop\prel\wpof{\aword}{\apropp}$.
\end{corollary}

%% file: appendix/proofs.tex

\section{Missing Proofs}

%
%
\begin{proof}[\Cref{thm:prop-rewriting,thm:hoare-characterizes-runs,thm:classification-equivalence}]
	The claims follow immediately from the definition.
\end{proof}

%
%
\begin{proof}[\Cref{thm:soundness-abstract-post}]
	Consider the interesting case $\aconf=(\astate, \amap) \stepofNEW{\atime}(\astatep,\amapp)=\aconfp$ of a non-stuttering step that is due to a transition of the form $\astate\transofNEW{\timeevent}{\aguard}{\anup}\astatep$ with $\anup = \setcompact{\vecof{x}\mapsto\vecof{y}}$ and $\amap \models \aform$ for some $(\astate, \aform) \in \aprop$.
	Then, $\amap + \atime \models \aformp \land \aguard$ where $\aformp$ corresponds to $\aform$ after waiting for $\atime$ time, $\aformp=\aform[\setclocks\mapsto\setclocks-\atime]\wedge\setclocks\geq\atime$.
	This means $\amap + \atime$ yields values for $\vecof{x}$ which can be chosen to satisfy $\amap+\atime\models \exists \vecof{z}. (\aformp \land \aguard)[\vecof{x} \mapsto \vecof{z}]$.
	The successor valuation $\amapp = \semof{\anup}(\amap+\atime)$ is constructed in such a way that it equals $\amap + \atime$ on all variables except $\vecof{x}$.
	On variables $x$ from $\vecof{x}$, it equals the assigned value $z$ from $\vecof{z}$, $\amapp(x) = z$.
	Similarly, for clock resets $\aclock$ in $\vecof{x}$ we have $\amapp(\aclock)=0$.
	We arrive at $\amapp \models \ssemof{\aguard}{\anup}{\atime}(\aform)$ and thus $(\astatep,\amapp)\models \postof{\atime}{\aprop}$.

	Assume $(\astatep,\amapp)\models\postof{\atime}{\aprop}$.
	In case of $\atime = 0$ and $(\astatep,\amapp)\models\aprop$, we are already done by a stuttering transition.
	Otherwise, there is a transition $\astate\transofNEW{\timeevent}{\aguard}{\anup}\astatep$ such that $(\astatep, \ssemof{\aguard}{\anup}{\atime}(\aform)) \in \postof{\atime}{\aprop}$ with $\amapp \models \ssemof{\aguard}{\anup}{\atime}(\aform)$.
	The existential quantifier guarantees the existence of a valuation $\amappp$ with $\amappp \models \aformp \land \aguard$ and $\amapp = \semof{\anup}(\amappp)$, where we use $\aformp=\aform[\setclocks\mapsto\setclocks-\atime]\wedge\setclocks\geq\atime$ as before.
	Valuation $\amapp$ evaluates variables and clocks in $\vecof{x}$ by the satisfying choices for the quantifier $\exists \vecof{z}$.
	Since $\aformp$ requires each clock value to be greater than $\atime$, there is $\amap$ with $\amap + \atime = \amappp$.
	Altogether, we conclude $(\astate, \amap) \stepofNEW{\atime}(\astatep,\amapp)$.
	
	The arguments are similar for transitions due to an event $\anevent$.
\end{proof}

%
%
\begin{proof}[\Cref{thm:sp-valid-hoare}]
	Assume $\spof{\aprop}{\aword} \prel \aprop$.
	We proceed by induction on $\aword$.
	In the base case, we have $\aword = \atime$.
	Consider some $\awordp = \atime_1.\ldots . \atime_n \equiv \atime$ and $\aconf \stepofNEW{\awordp}{} \aconfp$ with $\aconf \models \aprop$. 
	Using \Cref{thm:soundness-abstract-post} repeatedly, we get $\aconfp \models \spof{\aprop}{\aword}{}$.
	Thus, $\aconfp \models \apropp$.

	In the induction step, we have $\aword \equiv \awordp = \atime_1 \ldots \atime_n. \anevent. \awordpp$.
	Consider a run of the form $\aconf \stepofNEW{\atime_1 \ldots \atime_n}{} \aconfp \stepofNEW{\anevent}{} \aconfpp \stepofNEW{\awordp}{} \aconfppp$.
	Similar to the base case, $\aconfp \models \spost[\atime](\aprop) = \aprop'$, where $\atime =\atime_1 + \ldots + \atime_n$.
	\Cref{thm:soundness-abstract-post} now yields $\aconfpp \models \spost[\anevent](\aprop') = \aprop''$.
	By induction, we obtain the desired $\aconfppp \models \spof{\awordpp}{\aprop''}{} \prel \apropp$.
\end{proof}

%
%
\begin{proof}[\Cref{thm:soundness-abstract-pre}]
	Assume $\aconf = (\astate, \amap)$.
	Further, assume that for all $\aconfp$ we have $\aconf\stepofNEW{\asymbol}\aconfp$ implies $\aconfp \models \aprop$.
	To the contrary, assume ${\aconf = (\astate, \amap) \nmodels \preof{\asymbol}{\aprop} = \apropp}$.
	So there is \[
		(\astate,\smesof{\aguard}{\anup}{\atime}(\aformp)) \in \apropp
		\quad\text{and}\quad
		(\astatep, \aformp) \in \aprop
		\quad\text{and}\quad
		\astate \transofNEW{\asymbol}{\aguard}{\setcompact{\vecof{x} \mapsto \vecof{y}}}{}\astatep
	\]
	such that $\amap + \atime \nmodels \aformp[\vecof{x} \mapsto \vecof{y}]$ and $\amap + \atime \models \aguard$.
	We consider the case of a time progression $\asymbol = \Delta$, the remaining case of an event $\anevent$ is analogous.
	Since $\amap + \atime \models \aguard$, there is $\aconfp= (\astatep, \amapp)$ with $\aconf \stepofNEW{\atime} \aconfp$, namely $\amapp = \semof{\anup}(\amap + \atime)$.
	By assumption, $\aconfp \models \aprop$ and $\amapp \models \aformp$ (since we require only one symbolic configuration per state).
	But then $\amap + \atime \models \aformp[\vecof{x}\mapsto\vecof{y}]$:
	if $\amapp$ satisfies $\aformp$ then it also satisfies $\aformp[\vecof{x}\mapsto\vecof{y}]$ since the valuation of $\vecof{x}$ is unimportant.
	And since $\amapp$ coincides with $\amap+\atime$ on all variables and clocks outside $\vecof{x}$, also $\amap+\atime \models \aformp[\vecof{x}\mapsto\vecof{y}]$.

	Assume ${\aconf\models\preof{\asymbol}{\aprop}}$ with $\aconf = (\astate, \amap)$ and consider a step $\aconf \stepofNEW{\atime} \aconfp$ with $\aconfp = (\astatep, \amapp)$ due to transition $\astate \transofNEW{\Delta}{\aguard}{\anup}{} \astatep$.
	Then, $\amapp = \semof{\anup}(\amap + \atime)$.
	We have ${\amap \models \smesof{\aguard}{\anup}{\atime}(\aformp)}$ by assumption. 
	The condition $\lnot \aguard$ cannot be satisfied by $\amap + \atime$, due to the transition being enabled.
	But when $\amap + \atime \models \aformp[\vecof{x}\mapsto\vecof{y}]$, then changing the valuation of $\vecof{x}$ to $\vecof{y}$ satisfies $\aformp$.
	So, $\amapp \models \aformp$ and $\aconfp \models \aprop$.
\end{proof}

%
%
\begin{proof}[\Cref{thm:wp-valid-hoare}]
	Assume $\aprop\prel\wpof{\aword}{\apropp}{}$.
	We proceed by induction.
	In the base case $\aword = \atime \equiv \awordp = \atime_1 \ldots \atime_n$.
	Consider $\aconf \models \wpof{\aword}{\apropp}$ and $\aconf \stepofNEW{\awordp} \aconfp$.
	Then, by repeatedly applying \Cref{thm:soundness-abstract-pre} and the definition of $\narrowing$, we obtain $\aconfp \models \apropp$.
	
	In the induction step, we have that $\aword$ is equivalent to $\awordp.\anevent.\atime_1 \ldots \atime_n$.
	Consider a run of the form $\aconf \stepofNEW{\awordp} \aconfp \stepofNEW{\anevent} \aconfpp \stepofNEW{\atime_1ldots \atime_n}{} \aconfppp$.
	By induction, $\aconf \models \wpof{\aword}{\apropp'}{}$ implies $\aconfp \models \apropp' = \wpre[\anevent](\apropp'')$.
	Using \Cref{thm:wp-valid-hoare} yields $\aconfpp \models \apropp'' = \wpre[\atime](\apropp)$, where $\atime =\atime_1 + \ldots + \atime_n$.
	Similar to the base case, we obtain $\aconfp \models \apropp$.
\end{proof}

%
%
\begin{proof}[\Cref{thm:sp-characterizes-runs}]
	We show the following, slightly stronger proposition:
	for all $\aprop$ and $\aword$, if there is a sequence of steps on $\awordp \equiv \aword$ in $\anaut$ starting in some $\aconf \models \aprop$ (not necessarily initial) and ending in $\aconfp$, then $\aconfp \models \spof{\aprop}{\aword}\neq\false$.
	We proceed by induction.
	
	In the base case, consider $\aword = \atime$.
	Let $\awordp$ be a decomposition of $\aword$, $\aword\tequiv\awordp$.
	By definition, $\awordp$ is of the form $\awordp = \atime_1\dots\atime_n$ with $\atime = \atime_1 +\cdots+ \atime_n$.
	A run on $\awordp$ takes the form $\aconf=\aconf_0 \stepofNEW{\atime_1} \aconf_1 \stepofNEW{\atime_2} \cdots \stepofNEW{\atime_n} \aconf_n=\aconfp$.
	\Cref{thm:soundness-abstract-post} yields $\aconf_1\models\postof{\atime}{\aprop}$.
	Repeatedly applying \Cref{thm:soundness-abstract-post} thus gives $\aconf_n\models\post{\atime_n}\circ\cdots\circ\postof{\atime_1}{\aprop}$.
	By definition, this means $\aconfp\models\spof{\aprop}{\aword}$.

	For the induction step, consider $\aword=\awordp.\anevent.\atime$.
	Assume there is a step sequence from $\aconf \models \aprop$ on $\aword.\anevent.\atime$. 
	By induction, it visits $\aconfp \models \spof{\aprop}{\aword}{} = \apropp$ after $\aword$.
	\Cref{thm:soundness-abstract-post} yields that the next configuration is $\aconfpp \models \postof{\anevent}{\apropp}$.
	Finally, with same line of argument used in the base case we conclude that the sequence terminates in some configuration $\aconfppp\models\spof{\postof{\anevent}{\apropp}}{\atime}$.
	Hence, $\spof{\aprop}{\aword.\anevent.\atime}\neq\false$.
\end{proof}

%
%
\begin{proof}[\Cref{thm:precise-sp}]
	Assume $\models\hoare{\aprop}{\atime}{\apropp}$ with $\atime\in\RRplus$, the remaining cases are similar.
	Consider $\aconfp\models\spof{\aprop}{\atime}$.
	We get:
	\begin{align*}
		\aconfp~\models&~~\widening\bigl(\exists\atime_1\cdots,\atime_n.~\atime=\atime_1+\cdots+\atime_n \land \post{\atime_n}\circ\cdots\circ\post{\atime_1}(\aprop)\bigr)_{n\in\nat}
		\\
		=&~\bigcup_{n\in\nat}\bigl\{\exists\atime_1\cdots,\atime_n.~\atime=\atime_1+\cdots+\atime_n \land \post{\atime_n}\circ\cdots\circ\post{\atime_1}(\aprop)\bigr\}
	\end{align*}
	where the equality is due to the assumption of $\widening$ being precise.
	This means there is a decomposition $\atime=\atime_1+\cdots+\atime_k$ with $\aconfp\models\atime=\atime_1+\cdots+\atime_k \land \post{\atime_k}\circ\cdots\circ\post{\atime_1}(\aprop)$.
	It is easy to see that this means $\aconfp\models\post{\atime_k}\circ\cdots\circ\post{\atime_1}(\aprop)$.
	\Cref{thm:soundness-abstract-post} now yields some $\aconf\models\aprop$ with $\aconf\stepofNEW{\atime_1\dots\atime_k}\aconfp$.
	Hence, $\aconfp\models\apropp$ must hold due to the validity of $\hoare{\aprop}{\atime}{\apropp}$.
\end{proof}

%
%
\begin{proof}[\Cref{thm:precise-wp}]
	Analogous to the proof of \Cref{thm:precise-sp}.
\end{proof}

%% file: main.bbl
\begin{thebibliography}{10}
\providecommand{\url}[1]{\texttt{#1}}
\providecommand{\urlprefix}{URL }
\providecommand{\doi}[1]{https://doi.org/#1}

\bibitem{DBLP:conf/rex/AbadiL91}
Abadi, M., Lamport, L.: An old-fashioned recipe for real time. In: {REX}
  Workshop. {LNCS}, vol.~600, pp. 1--27. Springer (1991)

\bibitem{AD94}
Alur, R., Dill, D.L.: A theory of timed automata. TCS  \textbf{126}(2),
  183--235 (1994)

\bibitem{DBLP:conf/popl/BallNR03}
Ball, T., Naik, M., Rajamani, S.K.: From symptom to cause: localizing errors in
  counterexample traces. In: {POPL}. pp. 97--105. {ACM} (2003)

\bibitem{published}
Becker, M., Meyer, R., Runge, T., Schaefer, I., van~der Wall, S., Wolff, S.:
  Model-based fault classification for automotive software. In: {APLAS}.
  Lecture Notes in Computer Science, vol. 13658, pp. 110--131. Springer (2022)

\bibitem{MBTAutomotive08}
Bringmann, E., Kr{\"{a}}mer, A.: Model-based testing of automotive systems. In:
  {ICST}. pp. 485--493. {IEEE} (2008)

\bibitem{DBLP:conf/vmcai/ChristESW13}
Christ, J., Ermis, E., Sch{\"{a}}f, M., Wies, T.: Flow-sensitive fault
  localization. In: {VMCAI}. {LNCS}, vol.~7737, pp. 189--208. Springer (2013)

\bibitem{DBLP:conf/popl/CousotC77}
Cousot, P., Cousot, R.: Abstract interpretation: {A} unified lattice model for
  static analysis of programs by construction or approximation of fixpoints.
  In: {POPL}. pp. 238--252. {ACM} (1977)

\bibitem{DBLP:conf/popl/CousotH78}
Cousot, P., Halbwachs, N.: Automatic discovery of linear restraints among
  variables of a program. In: {POPL}. pp. 84--96. {ACM} Press (1978)

\bibitem{DBLP:journals/jsyml/Craig57}
Craig, W.: Linear reasoning. {A} new form of the herbrand-gentzen theorem. J.
  Symb. Log.  \textbf{22}(3),  250--268 (1957)

\bibitem{DBLP:conf/icse/DickinsonLP01}
Dickinson, W., Leon, D., Podgurski, A.: Finding failures by cluster analysis of
  execution profiles. In: {ICSE}. pp. 339--348. {IEEE} (2001)

\bibitem{DBLP:conf/sigsoft/DiGiuseppeJ12a}
DiGiuseppe, N., Jones, J.A.: Concept-based failure clustering. In: {SIGSOFT}
  {FSE}. p.~29. {ACM} (2012)

\bibitem{DBLP:conf/oopsla/DilligDLM13}
Dillig, I., Dillig, T., Li, B., McMillan, K.L.: Inductive invariant generation
  via abductive inference. In: {OOPSLA}. pp. 443--456. {ACM} (2013)

\bibitem{DBLP:conf/fm/FlanaganL01}
Flanagan, C., Leino, K.R.M.: Houdini, an annotation assistant for esc/java. In:
  {FME}. {LNCS}, vol.~2021, pp. 500--517. Springer (2001)

\bibitem{Dart05}
Godefroid, P., Klarlund, N., Sen, K.: {DART:} directed automated random
  testing. In: PLDI. pp. 213--223. {ACM} (2005)

\bibitem{DBLP:conf/issta/GolaghaLPI19}
Golagha, M., Lehnhoff, C., Pretschner, A., Ilmberger, H.: Failure clustering
  without coverage. In: {ISSTA}. pp. 134--145. {ACM} (2019)

\bibitem{DBLP:conf/icse/GolaghaPFN17}
Golagha, M., Pretschner, A., Fisch, D., Nagy, R.: Reducing failure analysis
  time: An industrial evaluation. In: {ICSE-SEIP}. pp. 293--302. {IEEE} (2017)

\bibitem{DBLP:conf/icse/GolaghaRMHP18}
Golagha, M., Raisuddin, A.M., Mittag, L., Hellhake, D., Pretschner, A.:
  Aletheia: a failure diagnosis toolchain. In: {ICSE} (Companion Volume). pp.
  13--16. {ACM} (2018)

\bibitem{DBLP:conf/tacas/Groce04}
Groce, A.: Error explanation with distance metrics. In: {TACAS}. {LNCS},
  vol.~2988, pp. 108--122. Springer (2004)

\bibitem{DBLP:conf/spin/GroceV03}
Groce, A., Visser, W.: What went wrong: Explaining counterexamples. In: {SPIN}.
  {LNCS}, vol.~2648, pp. 121--135. Springer (2003)

\bibitem{Guo06}
Guo, L., Roychoudhury, A., Wang, T.: Accurately choosing execution runs for
  software fault localization. In: {CC}. {LNCS}, vol.~3923, pp. 80--95.
  Springer (2006)

\bibitem{1702875}
Haase, V.: Real-time behavior of programs. {IEEE} Transactions on Software
  Engineering  \textbf{SE-7}(5),  494--501 (1981)

\bibitem{OperationalAbstraction03}
Harder, M., Mellen, J., Ernst, M.D.: Improving test suites via operational
  abstraction. In: {ICSE}. pp. 60--73. {IEEE} (2003)

\bibitem{DBLP:conf/tacas/HaslbeckN18}
Haslbeck, M.P.L., Nipkow, T.: Hoare logics for time bounds - {A} study in meta
  theory. In: {TACAS} {(1)}. {LNCS}, vol. 10805, pp. 155--171. Springer (2018)

\bibitem{DBLP:journals/fac/Hooman94}
Hooman, J.: Extending hoare logic to real-time. Formal Aspects Comput.
  \textbf{6}(6A),  801--826 (1994)

\bibitem{iso-uds}
{ISO}: {ISO 14229-1:2020 Road vehicles --- Unified diagnostic services (UDS)
  --- Part 1: Application layer}. Standard ISO 14229-1:2020, International
  Organization for Standardization, Geneva, CH (2020)

\bibitem{DBLP:conf/tacas/JinRS02}
Jin, H., Ravi, K., Somenzi, F.: Fate and free will in error traces. In:
  {TACAS}. {LNCS}, vol.~2280, pp. 445--459. Springer (2002)

\bibitem{DBLP:conf/issre/JordanHFP20}
Jordan, C.V., Hauer, F., Foth, P., Pretschner, A.: Time-series-based clustering
  for failure analysis in hardware-in-the-loop setups: An automotive case
  study. In: {ISSRE} Workshops. pp. 67--72. {IEEE} (2020)

\bibitem{DBLP:conf/pldi/JoseM11}
Jose, M., Majumdar, R.: Cause clue clauses: error localization using maximum
  satisfiability. In: {PLDI}. pp. 437--446. {ACM} (2011)

\bibitem{King76}
King, J.C.: Symbolic execution and program testing. {CACM}  \textbf{19}(7),
  385--394 (1976)

\bibitem{Zeller07}
Leitner, A., Oriol, M., Zeller, A., Ciupa, I., Meyer, B.: Efficient unit test
  case minimization. In: {ASE}. pp. 417--420. {ACM} (2007)

\bibitem{DBLP:conf/sigsoft/LiuH06}
Liu, C., Han, J.: Failure proximity: a fault localization-based approach. In:
  {SIGSOFT} {FSE}. pp. 46--56. {ACM} (2006)

\bibitem{Lyndon1959AnIT}
Lyndon, R.: An interpolation theorem in the predicate calculus. Pacific Journal
  of Mathematics  \textbf{9},  129--142 (1959)

\bibitem{DBLP:conf/cav/McMillan03}
McMillan, K.L.: Interpolation and sat-based model checking. In: {CAV}. {LNCS},
  vol.~2725, pp. 1--13. Springer (2003)

\bibitem{DBLP:reference/mc/McMillan18}
McMillan, K.L.: Interpolation and model checking. In: Handbook of Model
  Checking, pp. 421--446. Springer (2018)

\bibitem{DBLP:conf/tacas/MouraB08}
de~Moura, L.M., Bj{\o}rner, N.: {Z3:} an efficient {SMT} solver. In: {TACAS}.
  {LNCS}, vol.~4963, pp. 337--340. Springer (2008)

\bibitem{TestingMLSurvey2022}
Pan, R., Bagherzadeh, M., Ghaleb, T.A., Briand, L.C.: Test case selection and
  prioritization using machine learning: a systematic literature review. Empir.
  Softw. Eng.  \textbf{27}(2), ~29 (2022)

\bibitem{DBLP:conf/fase/PhamKRR17}
Pham, V., Khurana, S., Roy, S., Roychoudhury, A.: Bucketing failing tests via
  symbolic analysis. In: {FASE}. {LNCS}, vol. 10202, pp. 43--59. Springer
  (2017)

\bibitem{DBLP:conf/tap/PodelskiSW16}
Podelski, A., Sch{\"{a}}f, M., Wies, T.: Classifying bugs with interpolants.
  In: TAP@STAF. {LNCS}, vol.~9762, pp. 151--168. Springer (2016)

\bibitem{DBLP:conf/icse/PodgurskiLFMMSW03}
Podgurski, A., Leon, D., Francis, P., Masri, W., Minch, M., Sun, J., Wang, B.:
  Automated support for classifying software failure reports. In: {ICSE}. pp.
  465--477. {IEEE} (2003)

\bibitem{DBLP:conf/kbse/RenierisR03}
Renieris, M., Reiss, S.P.: Fault localization with nearest neighbor queries.
  In: {ASE}. pp. 30--39. {IEEE} (2003)

\bibitem{Siemens}
Rothermel, G., Harrold, M.J., Ostrin, J., Hong, C.: An empirical study of the
  effects of minimization on the fault detection capabilities of test suites.
  In: {ICSM}. pp. 34--43. {IEEE} (1998)

\bibitem{ASEBook2016}
Sch{\"{a}}uffele, J., Zurawka, T.: Automotive Software Engineering -
  Grundlagen, Prozesse, Methoden und Werkzeuge effizient einsetzen {(6.}
  Aufl.). Vieweg (2016)

\bibitem{DBLP:conf/rex/SchneiderBM91}
Schneider, F.B., Bloom, B., Marzullo, K.: Putting time into proof outlines. In:
  {REX} Workshop. {LNCS}, vol.~600, pp. 618--639. Springer (1991)

\bibitem{DBLP:conf/icse/Schwartz-Narbonne15}
Schwartz{-}Narbonne, D., Oh, C., Sch{\"{a}}f, M., Wies, T.: {VERMEER:} {A} tool
  for tracing and explaining faulty {C} programs. In: {ICSE} {(2)}. pp.
  737--740. {IEEE} (2015)

\bibitem{MBTSurvey2012}
Utting, M., Pretschner, A., Legeard, B.: A taxonomy of model-based testing
  approaches. Softw. Test. Verification Reliab.  \textbf{22}(5),  297--312
  (2012)

\bibitem{DBLP:conf/kbse/WangR05}
Wang, T., Roychoudhury, A.: Automated path generation for software fault
  localization. In: {ASE}. pp. 347--351. {ACM} (2005)

\bibitem{DBLP:journals/tse/WongGLAW16}
Wong, W.E., Gao, R., Li, Y., Abreu, R., Wotawa, F.: A survey on software fault
  localization. {IEEE} Trans. Software Eng.  \textbf{42}(8),  707--740 (2016)

\bibitem{WHLM}
Wong, W.E., Horgan, J.R., London, S., Mathur, A.P.: Effect of test set
  minimization on fault detection effectiveness. Softw. Pract. Exp.
  \textbf{28}(4),  347--369 (1998)

\bibitem{WHMP}
Wong, W.E., Horgan, J.R., Mathur, A.P., Pasquini, A.: Test set size
  minimization and fault detection effectiveness: {A} case study in a space
  application. J. Syst. Softw.  \textbf{48}(2),  79--89 (1999)

\bibitem{RegressionSurvey2012}
Yoo, S., Harman, M.: Regression testing minimization, selection and
  prioritization: a survey. Softw. Test. Verification Reliab.  \textbf{22}(2),
  67--120 (2012)

\bibitem{Clustering09}
Yoo, S., Harman, M., Tonella, P., Susi, A.: Clustering test cases to achieve
  effective and scalable prioritisation incorporating expert knowledge. In:
  {ISSTA}. pp. 201--212. {ACM} (2009)

\bibitem{DBLP:conf/sigsoft/Zeller02}
Zeller, A.: Isolating cause-effect chains from computer programs. In: {SIGSOFT}
  {FSE}. pp. 1--10. {ACM} (2002)

\bibitem{DBLP:conf/icse/ZhangGG06}
Zhang, X., Gupta, N., Gupta, R.: Locating faults through automated predicate
  switching. In: {ICSE}. pp. 272--281. {ACM} (2006)

\end{thebibliography}
